\documentclass{IEEEtran}
\usepackage{algorithmic}
\usepackage{graphicx}
\usepackage{cite}
\usepackage{amsmath,amssymb,amsfonts}
\usepackage{algorithmic}
\usepackage{graphicx}
\usepackage{textcomp}
\usepackage{xcolor}
\usepackage{algorithmic}
\usepackage{algorithm}
\usepackage{lineno,hyperref}
\usepackage{multirow}
\usepackage{epstopdf}
\usepackage{graphicx}
\usepackage{float}
\usepackage{caption}

\usepackage{hyperref}

\usepackage{color,array,amsthm}

\usepackage{graphicx}

\newtheorem{property}{Property}

\newtheorem{remark}{Remark}
\newtheorem{assumption}{Assumption}
\def\BibTeX{{\rm B\kern-.05em{\sc i\kern-.025em b}\kern-.08em
    T\kern-.1667em\lower.7ex\hbox{E}\kern-.125emX}}
\begin{document}

\title{Robust Control For Spacecraft Attitude Tracking Under Multiple Physical Limitations with Guaranteed Performance\\
}

\author{Jiakun Lei,
	Tao Meng,
	Weijia Wang, 
	Chengjin Yin, 
	Zhonghe Jin
	\thanks{Jiakun Lei,Ph.D. School of Aeronautics and Astronautics, Zhejiang University, Hangzhou, China, 310027, 12124010@zju.edu.cn}
	\thanks{Tao Meng, Prof., School of Aeronautics and Astronautics, Zhejiang University, Hangzhou, China, 310027, mengtao@zju.edu.cn}
	\thanks{Weijia Wang, Ph.D., School of Aeronautics and Astronautics, Zhejiang University, Hangzhou, China, 310027, 12024055@zju.edu.cn}
	\thanks{Chengjin Yin, M.S., School of Aeronautics and Astronautics, Zhejiang University, Hangzhou, China, 310027, 22124030@zju.edu.cn}
	\thanks{Zhonghe Jin, Prof., School of Aeronautics and Astronautics, Zhejiang University, Hangzhou, China, 310027, Zhejiang Key Laboratory of micro-nano-satellite, jinzh@zju.edu.cn}
}

\maketitle

\begin{abstract}
This paper considers the prescribed performance control (PPC) of spacecraft attitude tracking under multiple physical constraints, focusing on the robust issues. A novel Barrier Lyapunov function is proposed to realize the guaranteed-performance control under angular velocity constraint without singularity. Additionally, an adaptive strategy for the performance function is presented to soften the constraint and quickly re-stabilize the system after severe disturbances occur, providing strong robustness. Further, an auxiliary system is designed to handle the input saturation issue, incorporating the actuator limitation into the system. Based on the proposed structure, a backstepping controller is developed accordingly using a double-layer PPC framework. Numerical simulation results are presented to validate the proposed controller framework's efficiency and robustness.
\end{abstract}


\section{Introduction}
For a real aerospace engineering space mission, constraints, performance, and robustness are eternal topics, such as the physical limitation including angular velocity constraint and torque output limitation.
Besides, some performance requirements, including maximum settling time and steady-state control accuracy, are required to be satisfying, while the robustness of a system should be guaranteed.
Generally speaking, these requirements may be hard to satisfy simultaneously. Motivated by this issue, this paper investigates the possible way to satisfy all these requirements in attitude tracking problems while ensuring the system appears strong robustness for severe perturbation.

Many works related to this field have been discovered before.
As for the attitude control problem under angular velocity constraint, it has been widely discovered by many researchers due to its application value. In \cite{wie1995feedback}, a quaternion-based feedback control law is proposed with the consideration of angular rate limitation and torque limitation. Nevertheless, the proposed controller lacks strict proof of its stability. Based on the idea of potential field,  \cite{hu2013robust} presents a nonlinear controller to deal with the angular velocity constraint and the control saturation issue. In \cite{li2016adaptive}, by utilizing the backstepping methodology, the virtual control law is designed to be explicitly bounded using the hyperbolic tangent function, and the angular velocity limitation is tackled in this way. 
Similarly, the actuator output limitation is one of the most widely discussed constraints for the attitude control problem. Generally speaking, this issue is usually solved by introducing the auxiliary system, as introduced in \cite{shao2018fault,sun2017disturbance,zou2016finite}.

The attitude control problem with preassigned performance requirements has been of high-interest in recent years, and the prescribed performance control (PPC) scheme is often adopted to handle this problem, as stated in \cite{hu2018adaptive,zhang2019observer,wei2018learning,wei2021overview}. However, the traditional PPC scheme has the inherent risk of suffering from the singularity problem. As elaborated in \cite{bechlioulis_adaptive_2009}, a premise should be guaranteed for PPC such that the state constraint should stay in the constraint region at any time. Nevertheless, although this premise can be guaranteed technically at the initial condition, this premise may still not be able to satisfy when there exists significant external disturbance or input saturation \cite{SAPPC2022,yong2020flexible}.
This inherent problem significantly lowers the application value of traditional PPC in real engineering scenarios, and this issue has recently raised some attention, with several efficient solutions proposed. In \cite{SAPPC2022,WANG2022}, we propose two solutions individually, whose main idea is to loosen the original hard constraint into a soft one, ensuring that the PPC framework can work appropriately even when the constraint is violated. In \cite{yong2020flexible}, an auxiliary feedback system for the performance function is proposed to generate an additional signal, ensuring that the performance function's envelope can cover the state trajectory under any condition, and the singularity is circumvented in this way.

According to the existing work, for an efficient scheme fusing all these requirements, there is still a lack up to now. Although the PPC control under physical limitation has been investigated in \cite{golestani2022prescribed} recently and has provided a motivated solution, it still failed to handle the critical singularity problem. In this paper, we are dedicated to satisfying all these constraints simultaneously.
Compared with existing works, this paper delivers a possible framework to fuse the physical limitation, performance requirements, and robust characteristics together. The main contribution of this paper can be summarized as follows:

\textbf{1.} A newly-designed Barrier Lyapunov function is proposed to cope with the attitude tracking with preassigned performance requirements under angular velocity constraints. Further, the singularity problem is circumvented simultaneously.
$\textbf{2.}$ An adaptive strategy for performance function is designed, ensuring that the system is able to re-stabilize short after a sudden severe disturbance occurs, providing strong robustness for the system.
$\textbf{3. }$ Due to the consideration of many constraints and the singularity problem, the proposed scheme may be of higher application value compared with traditional PPC schemes.

The rest of this paper is organized as follows: the problem formulation and mathematical lemma introduction are delivered in section \ref{formulation} separately, while the main idea of the proposed solution is elaborated in section \ref{solution}. Controller derivation is proposed in \ref{derivation}, and the final simulation validation of the proposed scheme is detailed in section \ref{simulation}.

\section{Problem Formulation}
\label{formulation}
\subsection{Notations}
Following notations are defined for this paper.
$\boldsymbol{I}_n$ represents the $n \times n$ identity matrix, while $\|\cdot\|$ denotes the Euclidean norm of a vector or the induced norm of a matrix. 
The operator symbol $\boldsymbol{b}^{\times}$ denotes the $3 \times 3$ skew-symmetric matrix for vector cross manipulation, i.e. $\boldsymbol{b}^{\times}\boldsymbol{s} = \boldsymbol{b}\times\boldsymbol{s}$. Further, 
$\text{diag}\left(b_i\right)$ represents a diagonal matrix whose diagonal line is consisted by the components of the given vector $\boldsymbol{b}$, and $\text{vec}\left(b_{i}\right)$ denotes a column vector such that $\text{vec}\left(b_{i}\right) = \left[b_{1},b_{2},...b_{i}\right]^{\text{T}}$.

\subsection{Attitude System Modeling}
Considering the attitude error system of a rigid-body spacecraft, its kinematic and dynamic system expressed in the normalized attitude error quaternion is expressed as follows \cite{xiao2011adaptive}:
\begin{equation}
	\centering
	\label{errorsystem}
	\begin{aligned}
		\dot{\boldsymbol{q}}_{ev} &= \boldsymbol{F}_{e}\boldsymbol{\omega}_e\quad
		\dot{q}_{e0} = -\frac{1}{2}\boldsymbol{q}^{\text{T}}_{ev}\boldsymbol{\omega_{e}} \\
		\boldsymbol{J}\dot{\boldsymbol{\omega}}_e &=  \boldsymbol{M}_{0} + \boldsymbol{\tau} + \boldsymbol{d}
	\end{aligned}
\end{equation}
where the attitude error quaternion is denoted as $\boldsymbol{q}_{e} = \left[\boldsymbol{q}_{ev}^{\text{T}},q_{e0}\right]^{\text{T}}\in\mathbb{R}^{4}$. The vector part and the scalar part of the attitude error quaternion is expressed as $\boldsymbol{q}_{ev}$ and $q_{e0}$, receptively. 
The inertial matrix of the spacecraft expressed in the body-fixed frame is denoted as $\boldsymbol{J}\in\mathbb{R}^{3\times 3}$, where $\boldsymbol{J}$ is a symmetric positive-definite matrix all along. $\boldsymbol{\tau}\in\mathbb{R}^{3}$ represents the system's control input, while $\boldsymbol{d}\in\mathbb{R}^{3}$ denotes the lumped unknown external disturbances.
Denote the desired angular velocity expressed in the desired body-fixed frame $\mathfrak{R}_{d}$ as $\boldsymbol{\omega}_{d}$, the error angular velocity expressed in the current body-fixed frame can be expressed as $\boldsymbol{\omega}_
{e} = \boldsymbol{\omega}_{s} - \boldsymbol{C}_{e}\boldsymbol{\omega}_{d}$, where $\boldsymbol{\omega}_{s}$ represents the current body-fixed angular velocity with respect to the inertial frame, and
$\boldsymbol{C}_{e}$ denotes the transformation matrix from $\mathfrak{R}_{d}$ to $\mathfrak{R}_{b}$.
$\boldsymbol{M}_{0} = \boldsymbol{J}\boldsymbol{\omega}^{\times}_e\boldsymbol{C}_e\boldsymbol{\omega}_d 
- \boldsymbol{J}\boldsymbol{C}_e\dot{\boldsymbol{\omega}}_d
-\boldsymbol{\omega}_s^{\times}\boldsymbol{J}\boldsymbol{\omega}_s$ represents the lumped dynamical term, $\boldsymbol{F}_{e}$ represents the Jacobian matrix of attitude error kinematic expressed as $\boldsymbol{F}_{e} = \frac{1}{2}\left(q_{e0}\boldsymbol{I_{3}} + \boldsymbol{q}_{ev}^{\times}\right)$. Note that the following property will be hold such that $\|\boldsymbol{F}^{-1}_{e}\| = \frac{2}{|q_{e0}|}$.

For the following analysis, define a saturation variable as $\Delta\boldsymbol{\tau} = \boldsymbol{\tau} - \boldsymbol{u}$, where $\boldsymbol{u}$ represents the command control input derived by controller. 

\subsection{Assumptions}
For the synthesize of the proposed control scheme, these assumptions are made in this paper.
\begin{assumption}
	{\label{assump_J}}
	The inertial matrix $\boldsymbol{J}$ of the spacecraft is a known symmetric positive-definite matrix. Accordingly, we have:
	\begin{equation}
		\lambda (\boldsymbol{J})_{\text{min}}\boldsymbol{x}^{\text{T}}\boldsymbol{x} \le \boldsymbol{x}^{\text{T}}\boldsymbol{J}\boldsymbol{x} \le
		\lambda (\boldsymbol{J})_{\text{max}}\boldsymbol{x}^{\text{T}}\boldsymbol{x}
	\end{equation}
where $\lambda\left(\cdot\right)$ represents the corresponding eigen value.
\end{assumption} 

\begin{assumption}
	{\label{assump_Dis}}
	The external disturbance is unknown but bounded by a known constant, i.e., $\|\boldsymbol{d}\|$ $\le D_{m}$.
\end{assumption}

\section{Control Objective}
The primary purpose of this paper is to develop a relevant controller to ensure that the given physical constraints and performance requirements can be satisfied simultaneously.
Also, the ultimate boundedness of the closed-loop signals should be guaranteed. Further, another purpose of this paper is to ensure that the system can rapidly recover from significant sudden external disturbances.

\section{Problem Solution}
\label{solution}

\subsection{Error Transformation Procedure}
As stated in \cite{bechlioulis_robust_2008}, the performance function should be assigned according to the given performance requirements firstly. Defining a performance function vector as $\boldsymbol{\rho}\left(t\right) = \left[\rho_{1}\left(t\right),...,\rho_{i}\left(t\right)\right]^{\text{T}}\in\mathbb{R}^{3}\left(i = 1,2,3\right)$, considering arbitrary system state variable denoted as $\boldsymbol{e}(t) = \left[e_{1}(t),...e_{i}(t)\right]^{\text{T}}\in\mathbb{R}^{3}$, the state constraint for $e_{i}(t)$ can be expressed as follows:
\begin{equation}
	\label{constraintqe}
	-\rho_{i}\left(t\right) < e_{i}\left(t\right) < \rho_{i}\left(t\right)
\end{equation}
Further, defining a translated error variable corresponding to $\boldsymbol{e}(t)$ as $\boldsymbol{\varepsilon}_{}\left(t\right) = \left[\varepsilon_{1}\left(t\right),...\varepsilon_{i}\left(t\right)\right]^{\text{T}}\in\mathbb{R}^{3}\left(i = 1,2,3\right)$, and the $i$ th component of $\boldsymbol{\varepsilon}$ is defined as $\label{constraint}
	\varepsilon_{i}\left(t\right) = \frac{e_{i}\left(t\right)}{\rho_{i}\left(t\right)}$.
Accordingly, the expected state constraint \ref{constraintqe} can be transformed into an equivalently one expressed as $|\varepsilon_{i}\left(t\right)| < 1$. Since $|\varepsilon_{i}|<1$ will be hold if $\|\boldsymbol{\varepsilon}\|^{2} < 1$ is satisfied, thus we strengthen the original constraint into the stronger one.

\subsection{Barrier Lyapunov Function Design}
\label{BLFdesign}

Inspired by our previous work \cite{WANG2022}, to realize a non-singular PPC control under angular velocity limitations, we propose the following Barrier Lyapunov Function $V_{\text{B}}$ in this paper:
\begin{equation}
	V_B = \frac{k}{2}F\ln\left[\cosh\left(\boldsymbol{\varepsilon}^{\text{T}}\boldsymbol{\varepsilon}/F\right)\right]
\end{equation}
where $k > 0$ and $F > 0$ are the design parameters. 
Take the the gradient of the proposed BLF $V_{B}$ with respect to $\boldsymbol{\varepsilon}$, we have:
\begin{equation}
	\nabla_{\varepsilon}V_{B} = k\tanh\left(\boldsymbol{\varepsilon}^{\text{T}}\boldsymbol{\varepsilon}/F\right)\boldsymbol{\varepsilon}^{\text{T}}
\end{equation}
Compared with the mostly-applied logarithmic-type BLF expressed as $V_{l} = \frac{1}{2}\ln\left(\frac{1}{1-\|\boldsymbol{\varepsilon}\|^{2}}\right)$, notably, $\|\nabla_{\varepsilon}V_{l}\| \to \infty$ will be hold if $\|\boldsymbol{\varepsilon}\|^{2} \to 1$ is hold, while $\|\nabla_{\varepsilon}V_{B}\|$ will not tend to infinity under such a condition. Therefore, the gradient of the proposed BLF will increasing at a relatively slow rate when $\|\boldsymbol{\varepsilon}\|^{2} \to 1$, providing mild controller output.
Applying the proposed BLF will turn the original "hard" constraint into a soft one. However, the asymptotical convergence of $\boldsymbol{\varepsilon}$ is still guaranteed.
For the following analysis, considering a function $v\left(x\right) = \ln\left(\cosh\left(x\right)\right)\left(x\ge0\right)$,these following properties will be obtained.
\begin{property}
	\label{P1}
	$\frac{1}{2}x\tanh\left(x\right) \le \ln\left[\cosh\left(x\right)\right] \le x\tanh\left(x\right)$ will be hold for $\forall x\in\left[0,+\infty\right)$.
\end{property}
Define $g\left(x\right) = \ln\left[\cosh x\right] - \frac{1}{2}x\tanh x$, considering $\cosh x\cdot\dot{g}\left(x\right)$, the property will be easily obtained.
\begin{property}
	\label{P2}
	For $k>0$, there exists a constant $m$, $k_{0}<k$ such that $k\tanh\left(mx\right) \ge x$ will be satisfied on $x\in\left[0,+k_{0}\right)$ all along.
\end{property}
This property can be proved geometrically easily , here we omitted for brevity.
\subsection{Prescribed Performance Function Design}
\label{PF}

In this paper, the prescribed performance function (PPF) is designed to be a composite one, consisting of a nominal part and an adaptive-updated part.

\textbf{1. Nominal Part of PPF}

The nominal part of the PPF is designed the same as the one stated in \cite{SAPPC2022}, expressed as follows:
\begin{equation}
	\label{RPF}
	\rho_{n}(t) = \begin{cases}
		\rho_{e}(t)=\left(\rho_{e0}-\rho_{e\infty}\right)e^{-lt}+\rho_{e\infty}&  0 \le t<t_{1} \\ 
		\rho_{p}(t)={a_1t^2+a_2t+a_3}                                          &  t_1\le t<t_2\\
		\rho_{c}(t)=g_\infty                                                   &  t_2 \le t
	\end{cases}
\end{equation}
where $t_{2}>0$ stands for the preassigned settling time, $\rho_{e0}$, $\rho_{e\infty}$ denotes the initial value and the terminal value of the the exponential function part respectively, $g_{\infty}$ denotes the terminal value.  $a_1$, $a_2$, $a_3$ and $t_1$ are coefficients needs solving later, where $t_{1}$ represents the time instant such that $\rho_{e}\left(t_{1}\right) = \rho_{p}\left(t_{1}\right)$ is satisfied.
By indicating $t_{2}$, $\rho_{e0}$, $\rho_{e\infty}$, $g_{\infty}$ and $l$, the coefficient $t_{1}$, $a_{1}$, $a_{2}$, $a_{3}$ can be calculated according to the smoothness connection condition, expressed as follows:
\begin{equation}
	\label{RPFsolve}
	\begin{matrix}
		\left[\dfrac{k}{2} \left(t_2 - t_1\right)-1\right] \left(\rho_{e0}-\rho_{e\infty}\right)e^{-lt_1}-\rho_{e\infty}+g_\infty=0 \\
		a_1 = \left(\rho_{e0}-\rho_{e\infty}\right)e^{-lt_1}/2\left(t_1-t_2\right)\\
		a_2 = -2a_1t_2\\
		a_3 = g_\infty+a_1t_2^2
	\end{matrix}
\end{equation}$  $
Note that the selecting of the parameter should guarantee that a real-number solution exists for the equation (\ref{RPFsolve}).
\begin{remark}
	\label{remarktwo}
	One main characteristic of the introduced PPF is that by choosing appropriate parameters, $t_{1}$ can be set close enough to $t_{2}$ such that $t_{1} - t_{2} \to 0$, which vanishes the parabola curve part practically. This property will make sense in the following controller design section.
\end{remark}

\textbf{2. Adaptive Performance Function Strategy}

In order to alleviate the over-control problem when the system suffers from strong disturbances, an adaptive strategy for the performance function is proposed. Denote the adaptive part of PPF as $\Delta\boldsymbol{\rho}$. The adaptive strategy takes the following form:
\begin{equation}
	\Delta \dot{\boldsymbol{\rho}} = -K_{\rho}\Delta\boldsymbol{\rho} + K_{\tau} \text{vec}\left(|\tanh c_{\tau}\Delta\tau_{i}|\right)
\end{equation}
where $K_{\rho}$, $K_{\tau}$, $c_{\tau}$ are design parameters that needs indicating.
The primary mechanism of the proposed adaptive strategy for PPF can be elaborated as follows: When the input saturation happens,  $\Delta\boldsymbol{\rho}$ will be triggered by $|\tanh c_{\tau}\Delta\tau_{i}|$. Thus, it generates a bounded positive signal to $\rho_{i}$. This will establish a wider constraint boundary, significantly reducing the value of $\varepsilon$. Therefore, the over-control problem will be alleviated in this way.

Assume that the state trajectory is disturbed by external disturbances at a steady state, then the state trajectory will deviate from the original status. Notably, since $\rho_{i}$ has converged to a small value, thus $\varepsilon_{i}$ will have several orders of magnitude increasing, and $\varepsilon_{i} = e_{i}/\rho_{i}$ will be a big value under such a condition.
This will produce a tremendous control input for the system, causing intensive chattering of the state trajectory. This problem will be validated later in section \ref{simulation}.
\begin{remark}
	 Different from the work in \cite{yong2020flexible}, since the newly-designed BLF solves the singularity problem, the primary purpose of introducing the adaptive strategy for PPF is to alleviate the chattering and smooth the system's convergence behavior. Hence, it derives a simple adaptive law, and the state variable is unnecessarily fed to our scheme's adaptive strategy.  
\end{remark}
By combining the nominal part $\boldsymbol{\rho}_{n}$ and the adaptive part $\Delta\boldsymbol{\rho}$ of the PPF, the composite PPF $\boldsymbol{\rho}$ is defined as 
$\boldsymbol{\rho} = \boldsymbol{\rho}_{n} + \Delta\boldsymbol{\rho}$

\section{Control Law Derivation}
\label{derivation}
Based on the proposed BLF, a backstepping controller is developed
using a double-layered PPC structure. Simultaneously, an adaptive strategy of PPF and an auxiliary system are designed to cope with the over-control and input saturation issues respectively. The sketch map of the proposed controller is illustrated as in Figure \ref{fig_s}:

\begin{figure}[hbt!]
	\centering 
	\includegraphics[scale = 0.3]{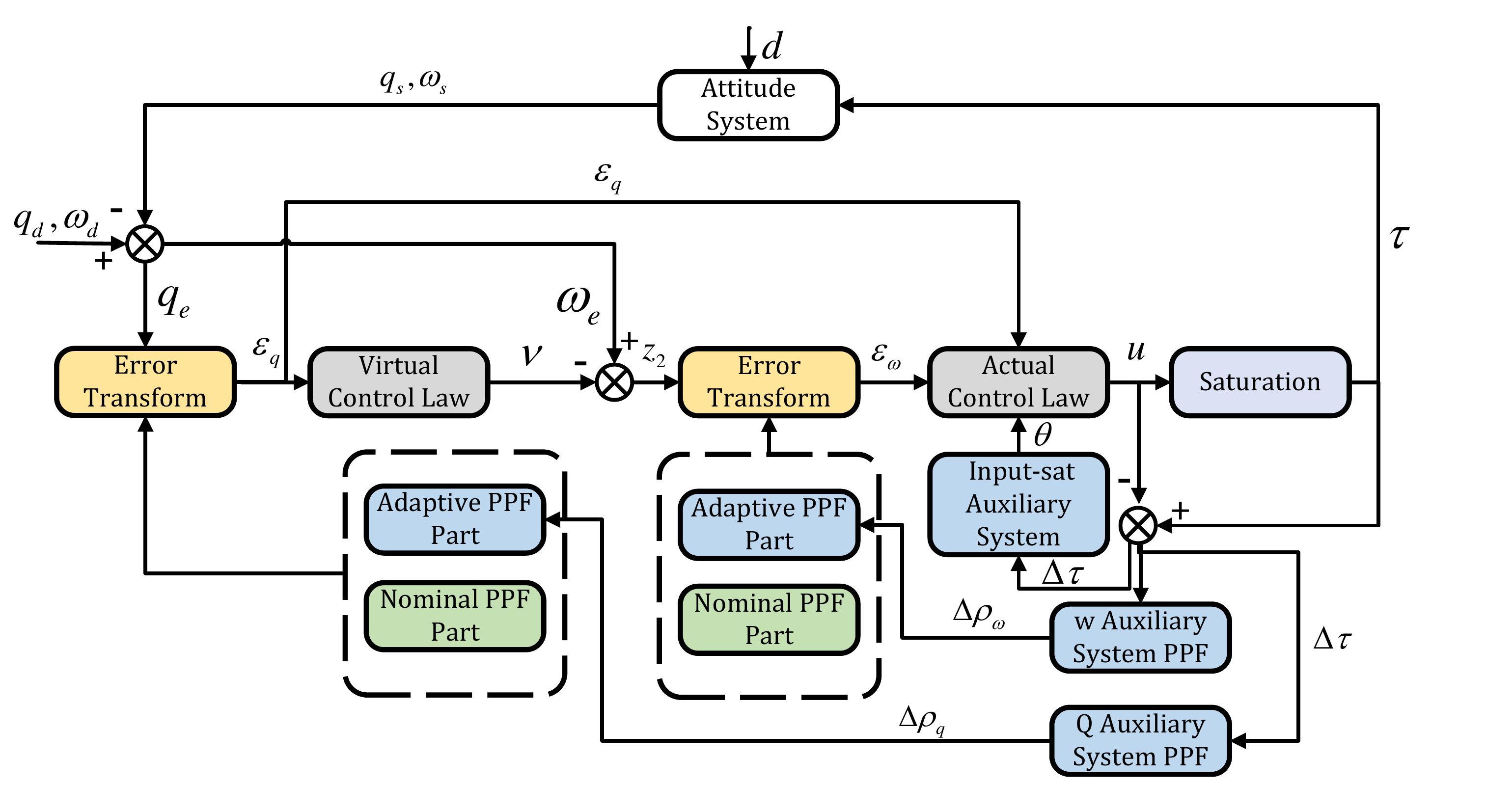}
	\caption{System Diagram}    
	\label{fig_s}  
\end{figure}

Define the PPF of $\boldsymbol{q}_{ev}$ as $\boldsymbol{\rho}_{q} = \boldsymbol{\rho}_{q0} + \Delta\boldsymbol{\rho}_{q}$, where $\boldsymbol{\rho}_{q0}$ and $\Delta\boldsymbol{\rho}_{q}$ denote the nominal part and the adaptive part correspondingly. Let $\rho_{qi}$ denotes the $i$ th component of the PPF for following analysis.
Define the corresponding transformed error variable of $\boldsymbol{q}_{ev}$ as $\boldsymbol{\varepsilon}_{q} = \left[\varepsilon_{q1},...\varepsilon_{qi}\right]^{\text{T}}\in\mathbb{R}^{3}$ such that $\varepsilon_{qi} = q_{evi}/\rho_{qi}$ is hold.
For further analyzing, considering the following two subsystem $\boldsymbol{z}_{1}$, $\boldsymbol{z}_{2}$ as:
\begin{equation}
	\boldsymbol{z}_{1} = \boldsymbol{\varepsilon}_{q}\quad
	\boldsymbol{z}_{2} = \boldsymbol{\omega}_{e} - \boldsymbol{v}
\end{equation}
where $\boldsymbol{v}$ denotes the virtual control law of $\boldsymbol{z}_{1}$ layer.

$\textbf{Step 1.}$ In view of the typical backstepping methodology, regard the attitude error angular velocity $\boldsymbol{\omega}_{e}$ as the virtual control law  $\boldsymbol{v}$. Take the time-derivative of $\boldsymbol{z}_{1}$, one has:
\begin{equation}
	\dot{\boldsymbol{z}}_{1} = \dot{\boldsymbol{\varepsilon}}_{q} = \boldsymbol{\psi}_{q}\boldsymbol{F}_{e}\boldsymbol{\omega}_{e} -\boldsymbol{\psi}_{q}\boldsymbol{\eta}_{q}\boldsymbol{q}_{ev}
\end{equation}
where $\boldsymbol{\psi}_{q} = \text{diag}\left(\psi_{q1},...\psi_{qi}\right)\in\mathbb{R}^{3\times 3}\left(i = 1,2,3\right)$ and $\boldsymbol{\eta}_{q} = \text{diag}\left(\eta_{q1},...\eta_{qi}\right)\in\mathbb{R}^{3\times 3}$ are diagonal matrices. $\psi_{qi}$ and $\eta_{qi}$ are defined as follows:
\begin{equation}
	\psi_{qi} = 1/\rho_{qi};\quad
	\eta_{qi} = \dot{\rho}_{qi}/\rho_{qi}
\end{equation}
Applying these notations,
the virtual control law $\boldsymbol{v}$ is designed and expressed as follows:
\begin{equation}
	\label{virtual}
	\boldsymbol{v} = -\frac{|q_{e0}|}{2}kM_{\omega}\boldsymbol{F}^{-1}_{e}\boldsymbol{\psi}_{q}^{-1}\text{vec}\left(\tanh\beta \varepsilon_{qi}\right)
\end{equation}
where $M_{\omega}>0$ is a design parameter, $\beta > 0$ is a big enough coefficient needs indicating, $k>1$ is a design parameter, $\text{vec}\left(\tanh\beta\varepsilon_{qi}\right)$ denotes the $\mathbb{R}^{3}$ element-spanned column vector.
\begin{remark}
	note $\|\boldsymbol{F}^{-1}_{e}\| = \frac{2}{|q_{e0}|}$, therefore the norm of the virtual control law satisfies:
$\|\boldsymbol{v}\| \le kM_{\omega}\frac{|q_{e0}|}{2}\|\boldsymbol{F}_{e}^{-1}\|\|\boldsymbol{\psi}_{q}^{-1}\|\le kM_{\omega}\left(|\rho_{qi}|\right)_{\text{max}}$. 
	Owing to the fact that $|\rho_{qi}|<=1$, it derives that $\|\boldsymbol{v}\| < kM_{\omega}$ will be always hold.
\end{remark}
Based on the proposed BLF, a candidate Lyapunov function $V_{1}$ is selected as follows:
\begin{equation}
	V_{1} = \frac{1}{2}k_{\text{1}}F_{1}\ln\left[\cosh\left(\boldsymbol{\varepsilon}_{q}^{\text{T}}\boldsymbol{\varepsilon}_{q}/F_{1}\right)\right]
\end{equation}
where $k_{1}$, $F_{1}$ are positive design parameters.
Take the time-derivative of $V_{1}$, one has:
\begin{equation}
	\begin{aligned}
		\dot{V}_{1} &=  k_{1}\tanh\left(\boldsymbol{\varepsilon}^{\text{T}}_{q}\boldsymbol{\varepsilon}_{q}/F_{1}\right)\boldsymbol{\varepsilon}^{\text{T}}_{q}\left[\boldsymbol{\psi}_{q}\boldsymbol{F}_{e}\boldsymbol{\omega}_{e} -\boldsymbol{\psi}_{q}\boldsymbol{\eta}_{q}\boldsymbol{q}_{ev}\right]	
	\end{aligned}
\end{equation}
Substituting the virtual control law (\ref{virtual}) into $\dot{V}_{1}$ yields:
\begin{equation}
	\label{dotV1}
	\begin{aligned}
		\dot{V}_{1} &= 
		-\frac{\|q_{e0}\|kM_{\omega}}{2}k_{1}\tanh\left(\boldsymbol{\varepsilon}^{\text{T}}_{q}\boldsymbol{\varepsilon}_{q}/F_{1}\right)\boldsymbol{\varepsilon}^{\text{T}}_{q}\text{vec}\left(\tanh\beta\varepsilon_{qi}\right)\\
		&\quad+k_{1}\tanh\left(\boldsymbol{\varepsilon}^{\text{T}}_{q}\boldsymbol{\varepsilon}_{q}/F_{1}\right)\boldsymbol{\varepsilon}^{\text{T}}_{q}\boldsymbol{\psi}_{q}\boldsymbol{F}_{e}\boldsymbol{z}_{2}\\
		&\quad -k_{1}\tanh\left(\boldsymbol{\varepsilon}^{\text{T}}_{q}\boldsymbol{\varepsilon}_{q}/F_{1}\right)\boldsymbol{\varepsilon}^{\text{T}}_{q}\boldsymbol{\psi}_{q}\boldsymbol{\eta}_{q}\boldsymbol{q}_{ev}	
	\end{aligned}
\end{equation}

Hence, 	according to property \ref{P2}, by selecting an appropriate $k$,
the first term of (\ref{dotV1}) can be rearranged into the following form: 
\begin{equation}
	\begin{aligned}
		&-\frac{\|q_{e0}\|kM_{\omega}}{2}k_{1}\tanh\left(\boldsymbol{\varepsilon^{\text{T}}_{q}\boldsymbol{\varepsilon_{q}}}/F_{1}\right)\boldsymbol{\varepsilon}^{\text{T}}_{q}\text{vec}\left(\tanh\beta\varepsilon_{qi}\right)\\
		\le& 			-\frac{\|q_{e0}\|M_{\omega}}{2}k_{1}\tanh\left(\boldsymbol{\varepsilon^{\text{T}}_{q}\boldsymbol{\varepsilon_{q}}}/F_{1}\right)\boldsymbol{\varepsilon}^{\text{T}}_{q}\boldsymbol{\varepsilon}_{q}
	\end{aligned}
\end{equation}
\begin{remark}
	\label{remarkboundeps}
Considering about the range of $|\varepsilon_{qi}|$. As we elaborated in Subsection \ref{BLFdesign}, owing to the designed adaptive system for PPF, the constraint boundary will getting wider if necessary, and this will sharply reduce the value of $|\varepsilon_{qi}|$ under extreme conditions. Therefore, it is rational to say $|\varepsilon_{qi}|$ is practically bounded. In this paper, we set $k=3$ for the following analysis.
\end{remark}
Subsequently, according to the property (\ref{P1}), note that $x\tanh x \ge \ln\left(\cosh x\right)$ is hold for $x > 0$. Hence, we have the final result expressed as:
\begin{equation}
			-\frac{\|q_{e0}\|M_{\omega}}{2}k_{1}\tanh\left(\boldsymbol{\varepsilon^{\text{T}}_{q}\boldsymbol{\varepsilon_{q}}}/F_{1}\right)\boldsymbol{\varepsilon}^{\text{T}}_{q}\boldsymbol{\varepsilon}_{q}
		\le  -\|q_{e0}\|M_{\omega}V_{1}
\end{equation}
Consider the third term in equation (\ref{dotV1}), owing to the fact that $\varepsilon_{qi} = q_{evi}/\rho_{qi}$, we can notice that $\boldsymbol{\psi}_{q}\boldsymbol{q}_{ev} = \boldsymbol{\varepsilon}_{q}$ will be always hold.
Further, according to the property (\ref{P1}), since $\frac{1}{2}x\tanh x \le \ln\cosh\left(x\right)$ is satisfied, thus we have the following result:
\begin{equation}
	\begin{aligned}
		&-k_{1}\tanh\left(\boldsymbol{\varepsilon}^{\text{T}}_{q}\boldsymbol{\varepsilon}_{q}/F_{1}\right)\boldsymbol{\varepsilon}^{\text{T}}_{q}\text{diag}\left(\frac{\dot{\rho}_{qi}}{\rho_{qi}}\right)\boldsymbol{\varepsilon}_{q}\\
		\le& 2\left(\frac{|\dot{\rho}_{qi}|}{\rho_{qi}}\right)_{\text{max}}\frac{k_{1}}{2}F_{1}\tanh\left(\boldsymbol{\varepsilon}^{\text{T}}_{q}\boldsymbol{\varepsilon}_{q}/F_{1}\right)\boldsymbol{\varepsilon}^{\text{T}}_{q}\boldsymbol{\varepsilon}_{q}/F_{1}\\
		\le & 4\left(\frac{|\dot{\rho}_{qi}|}{\rho_{qi}}\right)_{\text{max}}\frac{k_{1}}{2}F_{1}\ln\left[\cosh\left(\boldsymbol{\varepsilon}^{\text{T}}_{q}\boldsymbol{\varepsilon}_{q}/F_{1}\right)\right]
	\end{aligned}
\end{equation}
By sort out these results, we have the following conclusion:
\begin{equation}
	\begin{aligned}
		\dot{V}_{1} &\le -\left[\|q_{e0}\|M_{\omega} - 4\left(\frac{|\dot{\rho}_{qi}|}{\rho_{qi}}\right)_{\text{max}}\right]V_{1} \\
		&\quad +\tanh\left(\frac{\boldsymbol{\varepsilon}^{\text{T}}_{q}\boldsymbol{\varepsilon}_{q}}{F_{1}}\right)\boldsymbol{\varepsilon}^{\text{T}}_{q}\boldsymbol{\psi}_{q}\boldsymbol{F}_{e}\boldsymbol{z}_{2}
	\end{aligned}
\end{equation}

\textbf{Step 2.}
Considering the aforementioned error variable $\boldsymbol{z}_{2} = \boldsymbol{\omega}_{e} - \boldsymbol{v}$, take the time-derivative of $J\boldsymbol{z}_{2}$, one can be obtained that:
\begin{equation}
	\boldsymbol{J}\dot{\boldsymbol{z}}_{2} = \boldsymbol{M}_{0}+\boldsymbol{u}+\boldsymbol{d}+\Delta\boldsymbol{\tau}-\boldsymbol{J}\dot{\boldsymbol{v}}
\end{equation}
where $\boldsymbol{M}_{0}\in\mathbb{R}^{3}$ denotes the aforementioned dynamical terms, $\Delta\boldsymbol{\tau}$ represents the input saturation stated before. Similarly, the PPF of $\boldsymbol{z}_{2}$ is defined as $\boldsymbol{\rho}_{\omega} = \boldsymbol{\rho}_{\omega 0} + \Delta\boldsymbol{\rho}_{\omega}$, where the $i$ th component of $\boldsymbol{\rho}_{\omega}$ is expressed as $\rho_{\omega i}$.
Define the corresponding transformed error variable of $\boldsymbol{z}_{2}$ as $\boldsymbol{\varepsilon}_{\omega} = \left[\varepsilon_{\omega1},...\varepsilon_{\omega i}\right]^{\text{T}}\in\mathbb{R}^{3}$, such that $\varepsilon_{\omega} = z_{2i}/\rho_{\omega i}$ is hold, therefore the time derivative of $\boldsymbol{\varepsilon}_{\omega}$ can be expressed as:
\begin{equation}
	\label{domega}
	\dot{\boldsymbol{\varepsilon}}_{\omega} = \boldsymbol{\xi}\boldsymbol{J}^{-1}\cdot\boldsymbol{J}\dot{\boldsymbol{z}}_{2} -\boldsymbol{\xi} \boldsymbol{\gamma}\boldsymbol{z}_{2}
\end{equation}
where $\boldsymbol{\xi}$ is a $\mathbb{R}^{3 \times 3}$ diagonal matrix defined as $\boldsymbol{\xi} = \text{diag}\left(1/\rho_{\omega i}\right)\left(i=1,2,3\right)$, $\boldsymbol{\gamma}$ is a $\mathbb{R}^{3\times 3}$ diagonal matrix defined as $\boldsymbol{\gamma} = \text{diag}\left(\dot{\rho}_{\omega i}/\rho_{\omega i}\right)$.
For the $\boldsymbol{\omega}_{e}$, the actual command control law $\boldsymbol{u}$ is designed as follows:
\begin{equation}
	\begin{aligned}
			\label{controllaw}
		\boldsymbol{u} &= -\boldsymbol{M}_{0} - \hat{\boldsymbol{d}} + \boldsymbol{J}\dot{\boldsymbol{v}} - K_{\omega}\boldsymbol{J}\boldsymbol{\xi}^{-1}\boldsymbol{\varepsilon}_{\omega} 
		+ \boldsymbol{J}\boldsymbol{\gamma}\boldsymbol{z}_{2} - K_{u}\boldsymbol{J}\boldsymbol{\xi}^{-1}\boldsymbol{\theta} \\ &\quad-\frac{\tanh\left(\boldsymbol{\varepsilon}^{\text{T}}_{q}\boldsymbol{\varepsilon}_{q}/F_{1}\right)}{k_{2}\tanh\left(\boldsymbol{\varepsilon}^{\text{T}}_{\omega}\boldsymbol{\varepsilon}_{\omega}/F_{2}\right)} \boldsymbol{J}\boldsymbol{\xi}^{-1}\text{diag}\left(\rho_{\omega i}\right)\boldsymbol{\psi}_{q}\boldsymbol{F}_{e}\boldsymbol{\varepsilon}_{q}
	\end{aligned}
\end{equation}
where $\hat{\boldsymbol{d}} \in\mathbb{R}^{3}$ denotes the compensation for external disturbance expressed as $\hat{\boldsymbol{d}} = D_{m}\text{vec}\left(\tanh\frac{\varepsilon_{\omega i}}{\mu_{i}}\right)\left(\mu_{i}>0\right)$, $K_{\omega}$ is the controller gain to be designed. An auxiliary system $\boldsymbol{\theta}$ is designed to cope with the drawbacks brought by the input saturation issue, expressed as follows:
\begin{equation}
	\dot{\boldsymbol{\theta}} = -\left[K_{a}+ \frac{K_{b}\|\boldsymbol{\xi}\boldsymbol{J}^{-1}\Delta\boldsymbol{\tau}\|^{2}}{\|\boldsymbol{\theta}\|^{2}}\right]\boldsymbol{\theta} + \boldsymbol{\xi}\boldsymbol{J}^{-1}\text{vec}\left(\tanh\Delta\tau_{i}\right)
\end{equation}
where $K_{a}$, $K_{b}$ stands for the gain of the auxiliary system.
Specifically, based on the aforementioned idea in subsection \ref{PF}, the adaptive strategy for $\Delta\boldsymbol{\rho}_{q}$ and $\Delta\boldsymbol{\rho}_{\omega}$ are designed as follows:
\begin{equation}
	\begin{aligned}
		\Delta\dot{\boldsymbol{\rho}}_{q} &= -C_{q}\Delta\boldsymbol{\rho}_{q} + C_{\tau}\boldsymbol{\xi}\boldsymbol{J}^{-1}\text{vec}\left(|\tanh\Delta\tau_{i}|\right)\\
		\Delta\dot{\boldsymbol{\rho}}_{\omega} &= -C_{\omega}\Delta\boldsymbol{\rho}_{\omega} + B_{\tau}\boldsymbol{\xi}\boldsymbol{J}^{-1}\text{vec}\left(|\tanh\Delta\tau_{i}|\right)
	\end{aligned}
\end{equation}
where $C_{q}$, $C_{\omega}$, $C_{\tau}$, $B_\tau$ are gain parameters. By adjusting these parameter, the equilibrium value of the adaptive system is able to be changed arbitrarily. Notably, since the value of $|q_{evi}|$ will never larger than 1, thus the equilibrium value of PPF adaptive strategy should be set accordingly.
Choose these candidate Lyapunov functions as follows:
	$V_{2} =  \frac{k_{2}}{2}F_{2}\ln\left[\cosh
	\left(\boldsymbol{\varepsilon}^{\text{T}}_{\omega}\boldsymbol{\varepsilon}_{\omega}/F_{2}\right)\right]$, $V_{3} =   \frac{1}{2}\boldsymbol{\theta}^{\text{T}}\boldsymbol{\theta}$, $V_{4} = \frac{1}{2}\Delta\boldsymbol{\rho}^{\text{T}}_{q}\boldsymbol{\rho}_{q} + \frac{1}{2}\Delta\boldsymbol{\rho}^{\text{T}}_{\omega}\boldsymbol{\rho}_{\omega}$
Take the time-derivative of $V_{2}$, one can be obtained that:
\begin{equation}
	\label{dV2}
	\dot{V}_{2} = k_{2}\tanh\left(\boldsymbol{\varepsilon}_{\omega}^{\text{T}}\boldsymbol{\varepsilon}_{\omega}/F_{2}\right)\boldsymbol{\varepsilon}_{\omega}^{\text{T}}\dot{\boldsymbol{\varepsilon}}_{\omega}
\end{equation}

Considering the term expressed as follows, we have:
\begin{equation}
	\begin{aligned}\label{relationship}
		&\quad k_{2}\tanh\left(\boldsymbol{\varepsilon}^{\text{T}}_{\omega}\boldsymbol{\varepsilon}_{\omega}/F_{2}\right)\boldsymbol{\varepsilon}^{\text{T}}_{\omega}\boldsymbol{\xi}\boldsymbol{J}^{-1}\tilde{\boldsymbol{d}}\\
		& \le
		k_{2}\tanh\left(\boldsymbol{\varepsilon}^{\text{T}}_{\omega}\boldsymbol{\varepsilon}_{\omega}/F_{2}\right)\frac{\lambda\left(\boldsymbol{\xi}\right)_{\text{max}}}{\lambda\left(\boldsymbol{J}\right)_{\text{min}}}D_{m}\left[|\varepsilon_{\omega i}| - \varepsilon_{\omega i}\tanh\left(\frac{\varepsilon_{\omega i}}{\mu_{i}}\right)\right] \\
		&\le k_{2}\frac{\lambda\left(\boldsymbol{\xi}\right)_{\text{max}}}{\lambda\left(\boldsymbol{J}\right)_{\text{min}}}0.2785D_{\text{m}}\sum_{i=1}^{3}\mu_{i}
	\end{aligned}
\end{equation}

Substituting the actual control law into the equation (\ref{domega}) and combined with (\ref{dV2}).
Accordingly, define $\hat{D} = k_{2}\frac{\lambda\left(\boldsymbol{\xi}\right)_{\text{max}}}{\lambda\left(\boldsymbol{J}\right)_{\text{min}}}	0.2785D_{\text{m}}\sum_{i=1}^{3}\mu_{i}$, $P = k_{2}\tanh\left(\boldsymbol{\varepsilon}_{\omega}^{\text{T}}\boldsymbol{\varepsilon}_{\omega}/F_{2}\right)$, $R = \tanh\left(\frac{\boldsymbol{\varepsilon}^{\text{T}}_{q}\boldsymbol{\varepsilon}_{q}}{F_{1}}\right)\boldsymbol{\varepsilon}^{\text{T}}_{q}\boldsymbol{\psi}\boldsymbol{F}_{e}\boldsymbol{z}_{2}$ the equation can be rearranged as:
\begin{equation}
	\begin{aligned}
				\dot{V}_{2} &\le -K_{\omega}P\|\boldsymbol{\varepsilon}_{\omega}\|^{2} + K_{u}P\left(\frac{\|\boldsymbol{\varepsilon}_{\omega}\|^{2}}{2} + \frac{\|\boldsymbol{\theta}\|^{2}}{2}\right)
		\\&\quad+ P\left(\frac{\|\boldsymbol{\varepsilon}_{\omega}\|^{2}}{2} + \frac{\|\boldsymbol{\xi}\boldsymbol{J}^{-1}\Delta\boldsymbol{\tau}\|^{2}}{2}\right) - R + \hat{D}
	\end{aligned}
\end{equation}
In view of the fact in property (\ref{P2}), it should be noted that $PF_{2}\left[\|\boldsymbol{\varepsilon}_{\omega}\|^{2}/F_{2}\right] \ge k_{2}F_{2}\ln\left[\cosh\left(\boldsymbol{\varepsilon}_{\omega}^{\text{T}}\boldsymbol{\varepsilon}_{\omega}/F_{2}\right)\right] = 2V_{2} \ge \frac{1}{2}PF_{2}\left[\|\boldsymbol{\varepsilon}_{\omega}\|^{2}/F_{2}\right]$. Hence, the inequality can be rearranged into the following form:
\begin{equation}\begin{aligned}
		\dot{V}_{2} &\le -2K_{\omega}V_{2} + 2K_{u}V_{2} + 2V_{2} + \frac{K_{u}k_{2}}{2}\|\boldsymbol{\theta}\|^{2}
		\\&\quad+\frac{k_{2}}{2}\|\boldsymbol{\xi}\boldsymbol{J}^{-1}\Delta\boldsymbol{\tau}\|^{2} + \hat{D} - R
	\end{aligned}
\end{equation}

Take the time-derivative of $V_{3}$, one can be obtained that:
\begin{equation}
	\dot{V}_{3} 
	\le -\left(K_{a} - \frac{1}{2}\right)\|\boldsymbol{\theta}\|^{2} -\left(K_{b} - \frac{1}{2}\right)\|\boldsymbol{\xi}\boldsymbol{J}^{-1}\Delta\boldsymbol{\tau}\|^{2}
\end{equation}
Take the time-derivative of $V_{4}$, one can be obtained that:
\begin{equation}
	\label{dV4}
	\begin{aligned}
		\dot{V}_{4} &= -C_{q}\|\Delta\boldsymbol{\rho}_{q}\|^{2} - C_{\omega}\|\Delta\boldsymbol{\rho}_{\omega}\|^{2}\\&\quad +C_{\tau}\Delta\boldsymbol{\rho}^{\text{T}}_{q}\boldsymbol{\xi}\boldsymbol{J}^{-1} \text{vec}\left(|\tanh\Delta\tau_{i}|\right)\\&\quad +B_{\tau}\Delta\boldsymbol{\rho}^{\text{T}}_{\omega}\boldsymbol{\xi}\boldsymbol{J}^{-1}\text{vec}\left(|\tanh\Delta\tau_{i}|\right)
	\end{aligned}
\end{equation}
Applying the Young's inequality, the equation (\ref{dV4}) can be further written as:
\begin{equation}
	\begin{aligned}
			\dot{V}_{4} &\le -\frac{1}{2}\left(2C_{q} - C_{\tau}\right)\|\Delta\boldsymbol{\rho}_{q}\|^{2}  
		- \frac{1}{2}\left(2C_{\omega} - B_{\tau}\right)\|\Delta\boldsymbol{\rho}_{\omega}\|^{2}
		\\&\quad+ \frac{1}{2}\left[C_{\tau}+B_{\tau}\right]\|\boldsymbol{\xi}\boldsymbol{J}^{-1}\Delta\boldsymbol{\tau}\|^{2}
	\end{aligned}
\end{equation}

Define a composite Lyapunov function as $V = V_{1}+V_{2}+{V}_{3}+V_{4}$, take the time-derivative of $V$ and sort out all these results, we have the following conclusion:
\begin{equation}
	\begin{aligned}
		\dot{V} \le -S_{1}V_{1}-S_{2}V_{2}-S_{3}V_{3}-S_{4}V_{4}+\hat{D}
	\end{aligned}
\end{equation}
where $S_{1} = \|q_{e0}\|M_{\omega} - 4\left(\frac{|\dot{\rho}_{i}|}{\rho_{i}}\right)_{\text{max}}$, $S_{2} = 2K_{\omega}-2K_{u}-2$, $S_{3} = 2K_{a}-1-K_{u}k_{2}$, $S_{4} = \min\left(2C_{q}-C_{\tau},2C_{\omega}-B_{\tau}\right)$. The main principle of the parameter selecting is $S_{1},S_{2},S_{3},S_{4}$ should be guaranteed to be positive. Further, $2K_{b}-1-k_{2}-C_{\tau}-B_{\tau} < 0$ should also be guaranteed to be satisfied. In view of these principle, it derives the following result:
\begin{equation}
	\dot{V} \le -\min\left(S_{1},S_{2},S_{3},S_{4}\right)V + \hat{D}
\end{equation}
In view of the final result, the ultimately boundedness of the system is ensured, the system will finally converge to a residual set. The state trajectory of $\boldsymbol{q}_{ev}$ and $\boldsymbol{z}_{2}$ will converge to the desired performance constraint region, and all these performance requirements will be able to be satisfied.
Here we give main principles for the parameter selecting.

\textbf{1.} To ensure that $\boldsymbol{z}_{1}$ is able to converge, it should be guaranteed that $\|q_{e0}\|M_{\omega} - 4\left(\frac{|\dot{\rho}_{qi}|}{\rho_{qi}}\right)_{
	\text{max}} >0$ is always hold. 
Considering about this term expressed as $\frac{|\dot{\rho}_{qi}|}{\rho_{qi}}$. As we stated in Remark \ref{remarktwo}, $t_{1}-t_{2}$ can be set close enough to 0 practically, this ensures that the $\rho_{qi}$ can be approximately regarded as an exponential-decayed one for the whole convergence stage. Hence, $\left(\frac{|\dot{\rho}_{qi}|}{\rho_{qi}}\right)_{\text{max}} \approx l$ is satisfied in this way.
\textbf{2. } Considering the last term in the control law (\ref{controllaw}), a small enough positive constant $\sigma$ can be added to the denominator as  $k_{2}\tanh\left(\boldsymbol{\varepsilon}^{\text{T}}_{q}\boldsymbol{\varepsilon}_{q}/F_{2}\right) + \sigma$ to avoid the potential singularity problem.

\section{Numerical Simulation}
\label{simulation}
In this section, an assumed attitude tracking task is established, with specific requirements and constraints are presented for this virtual space mission. Simulation results are illustrated as below for the validation of the proposed scheme.
The spacecraft is assumed to be a rigid-body one, of which the inertial matrix expressed in the body-fixed frame is expressed as $\boldsymbol{J} =  \text{diag}\left(2.8,2.5,1.9\right)Kg\cdot m^{2}$. The external disturbance is supposed to be a time-varying one expressed as follows:
\begin{equation}
	\boldsymbol{d} = 
	\begin{bmatrix}
		1e-4 \cdot \left[4\sin\left(3\omega_{p}t\right) + 3\cos\left(10\omega_{p}t\right) -20\right]\\ 	
		1e-4\left[-1.5\sin\left(2\omega_{p}t\right) + 3\cos\left(5\omega_{p}t\right) +20\right]\\ 	
		1e-4\left[3\sin\left(10\omega_{p}t\right) - 8\cos\left(4\omega_{p}t\right) +20\right]\\ 	
	\end{bmatrix}
\end{equation}
where $\omega_{p} = 0.01$.
Further, the desired attitude quaternion $\boldsymbol{q}_{d}$ and the desired attitude angular velocity $\boldsymbol{\omega}_{d}$ is set as follows:
\begin{equation}
	\begin{aligned}
		\boldsymbol{q}_{d}\left(0\right) &= \left[0.2,-0.5,-0.5,-0.6782\right]^{\text{T}}\\
		\boldsymbol{\omega}_{d}\left(t\right) &= 0.5 [\cos\left(t/30\right),\sin\left(t/20\right),-\cos\left(t/40\right)]^{\text{T}}
	\end{aligned}
\end{equation}

The constraint for the attitude tracking task is state as follows.

\textbf{Physical Constraints}. Physical constraints for the attitude tracking is elaborated as follows: 
\textbf{1.} The maximum rotation speed should not exceed $3^{\circ}/s$. \textbf{2.} The maximum output on each axis is assumed to be symmetric such that $\tau_{m} = 0.05Nm$, while the minima of controller output is assumed to be $1e-4Nm$.

\textbf{Performance Requirements}. Performance requirements of the assumed attitude tracking task is stated as follows: \textbf{1.} The tracking error should converge to no more than $\left(|q_{evi}|\right)_{\text{max}} < 5e-3$ in no more than $60s$. $\textbf{2.}$ The terminal control error should satisfy $|q_{evi}|_{\text{max}} < 5e-4$, which is corresponding to $0.03^{\circ}$.

\subsection{Normal Case Attitude Tracking Simulation}
The initial condition of the spacecraft is randomly chosen as follows:
\begin{equation}
	\begin{aligned}
		\boldsymbol{q}_{s}\left(0\right) &= \left[0.1554,0.4271,0.4792,0.7509\right]^{\text{T}}\\
		\boldsymbol{\omega}_{s}\left(0\right) &= \left[0,0,0\right]^{\text{T}}
	\end{aligned}
\end{equation}

\begin{figure}[hbt!]
	\centering 
	\includegraphics[scale = 0.52]{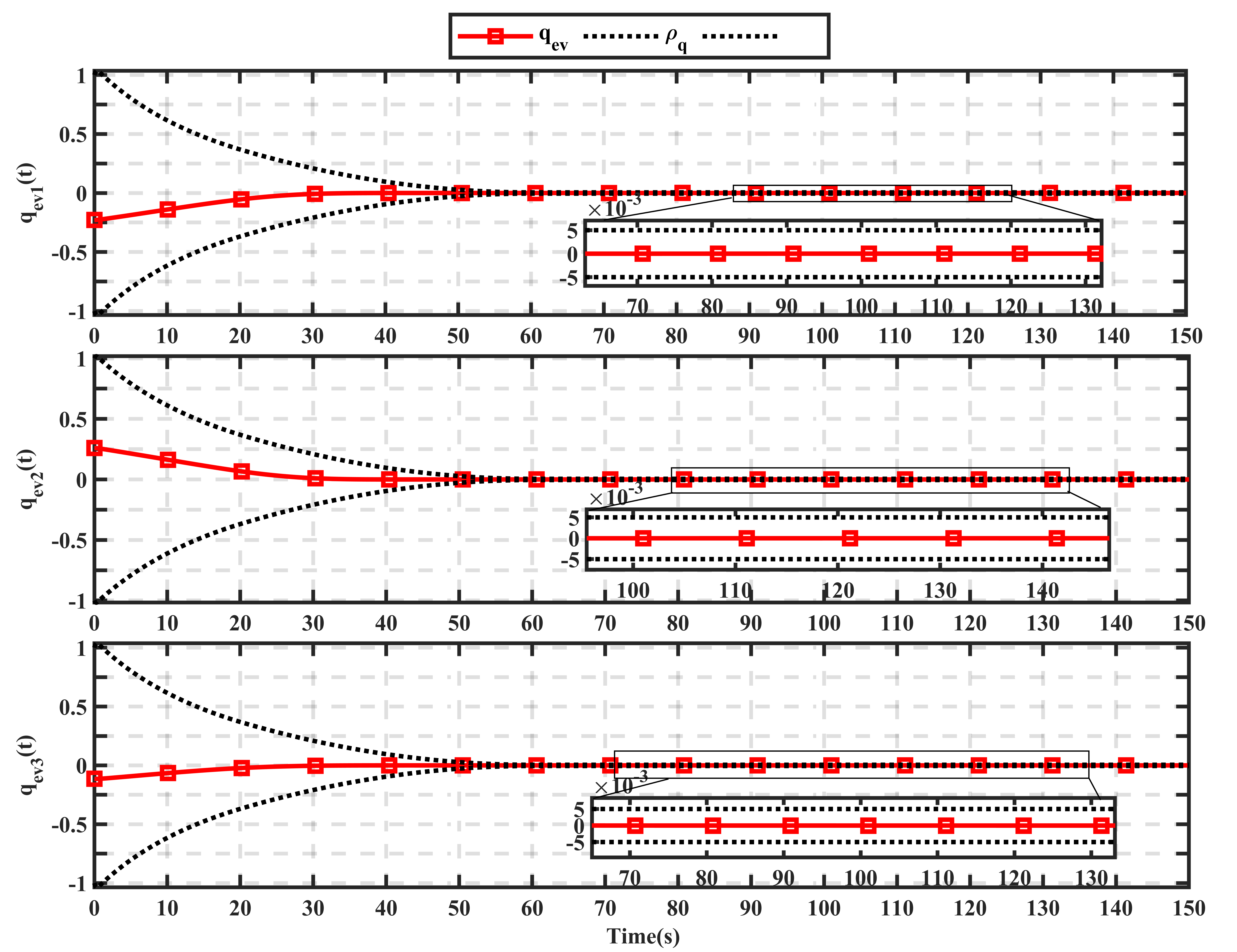}
	\caption{Time Responding of $\boldsymbol{q}_{ev}$ (Normal Case)}      
	\label{fig_qe_normal}   
	\centering 
	\includegraphics[scale = 0.46]{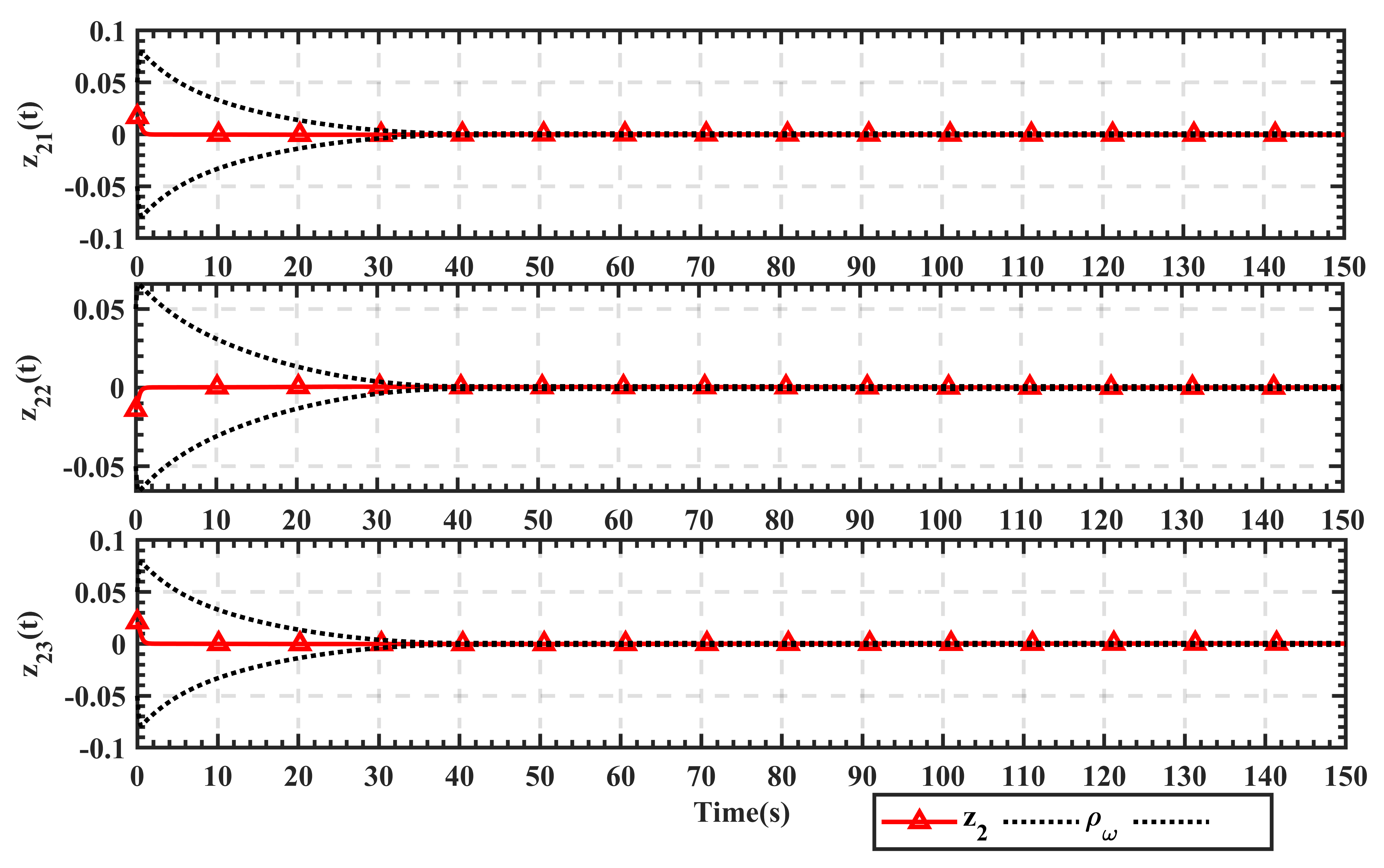}
	\caption{Time Responding of $\boldsymbol{z}_{2}$ (Normal Case)} 
	\label{fig_z_normal}   
	\centering 
	\includegraphics[scale = 0.465]{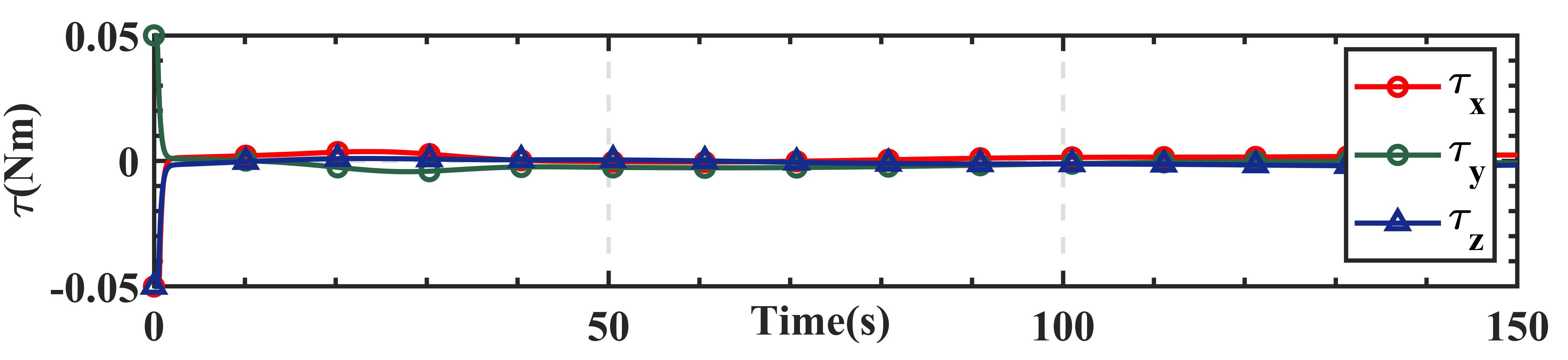}
	\caption{Time Responding of $\boldsymbol{\tau}$ (Normal Case)}    
	\label{fig_u_normal} 
		\centering 
	\includegraphics[scale = 0.5]{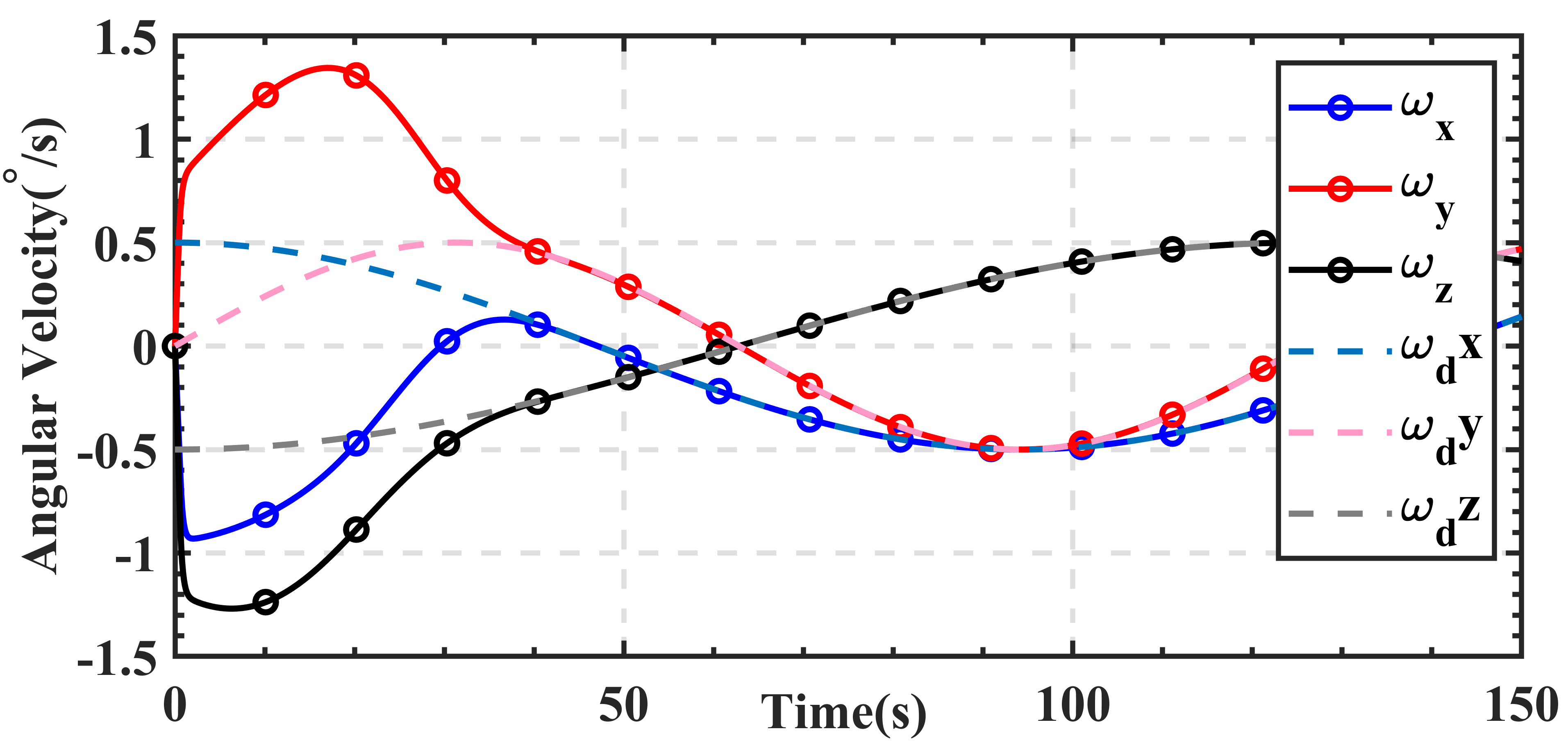}
	\caption{Time Responding of $\boldsymbol{\omega}_{s}$ (Normal Case)}    
	\label{fig_w_normal}
\end{figure}

Considering the fact that the maxima of the desired tracking angular velocity is $0.5\cdot\frac{\sqrt{3}}{3}^{\circ}/s$, we set $k=3$, $M_{\omega} = 0.017 = 1 \cdot \frac{\pi}{180}$ for safety consideration in this section. It should be noticed that the given desired angular velocity should not exceed the expected rotation rates limitation.
Further, according to the performance requirements, the nominal PPF for $\boldsymbol{q}_{e}$ layer is designed in Table [\ref{tab:RPFpara1}]:

\begin{table}[hbt!]
	\centering
	\begin{tabular}{|l|l|}
		\hline
		Initial Value of Exponential Function Part $\rho_{e0}$  &  1\\ \hline
		Terminal Value of Exponential Function Part $\rho_{e\infty}$ &  1e-4 \\ \hline 
		Coefficient of Exponential Function Part $l$ & 0.05 \\ \hline
		Convergence Time of the RPF   $t_{2}$              & 60 \\ \hline
		Terminal Value of the RPF   $g_{\infty}$         & 5e-3 \\ \hline
	\end{tabular}
	\caption{\label{tab:RPFpara1} Coefficients for the designed RPF $\rho_{q}$}
\end{table}

Accordingly, to guarantee that the $\boldsymbol{z}_{2}$ subsystem is able to converge in a short time, the nominal PPF for $\boldsymbol{z}_{2}$ layer is designed in Table [\ref{tab:RPFpara2}]:
\begin{table}[hbt!]
	\centering
	\begin{tabular}{|l|l|}
		\hline
		Initial Value of Exponential Function Part $\rho_{e0}$  &  0.08\\ \hline
		Terminal Value of Exponential Function Part $\rho_{e\infty}$ &  1e-6 \\ \hline 
		Coefficient of Exponential Function Part $l$ & 0.5 \\ \hline
		Convergence Time of the RPF   $t_{2}$              & 40 \\ \hline
		Terminal Value of the RPF   $g_{\infty}$         & 3e-5 \\ \hline
	\end{tabular}
	\caption{\label{tab:RPFpara2} Coefficients for the designed RPF for $\rho_{\omega}$}
\end{table}

In Figure[\ref{fig_qe_normal}] and Figure [\ref{fig_z_normal}], it can be observed that both $\boldsymbol{q}_{ev}$ and $\boldsymbol{z}_{2}$ is able to rapidly converge to the constraint region, and the performance requirements are satisfied. The terminal control error is bounded by $1e-4$ which indicates that the given performance requirements are satisfied. It can be observed in Figure [\ref{fig_w_normal}] that the maxima of the spacecraft's rotation speed is 2.1$deg/s$, which is below the given limitation $3^{\circ}/s$. Also, the actual angular velocity rapidly converges to the desired signals, which is illustrated in dotted lines. The simulation result indicates that all the physical limitations, performance constraint are satisfied simultaneously.

\subsection{Validation of Robustness}

In this subsection, severe sudden disturbance will be exerted on the system in order to validate the robustness of the proposed scheme. An additional significant sudden external will be exerted to the system at $t = 20s$ and $t = 80s$. The proposed scheme is expected to re-stabilizing the spacecraft in a short time after the sudden impact.

\begin{figure}[hbt]
	\centering 
	\includegraphics[scale = 0.5]{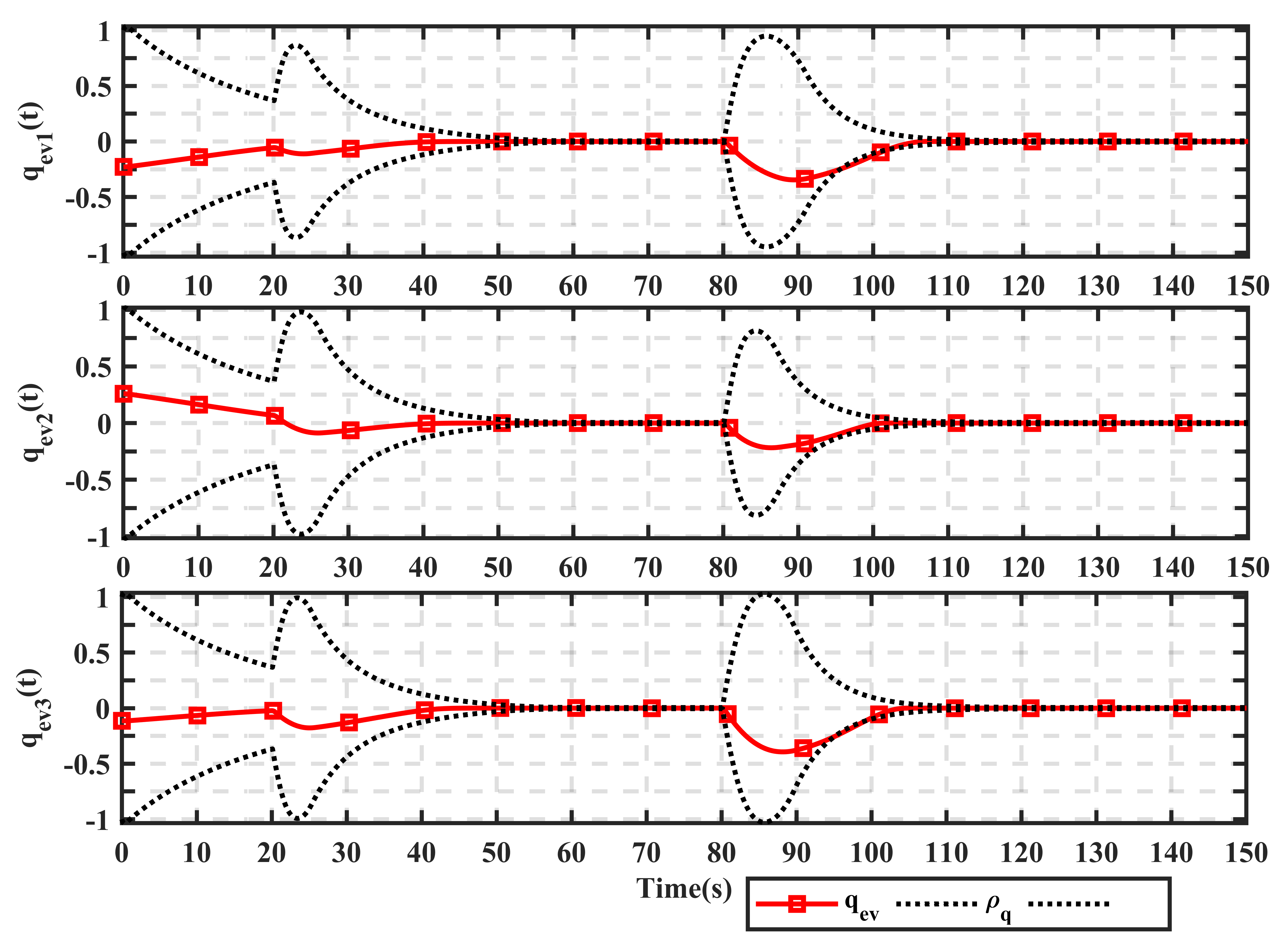}
	\caption{Time Responding of $\boldsymbol{q}_{ev}$ (Sudden Disturbance Case)}  
	\label{fig_qe_dis}    
	\centering 
	\includegraphics[scale = 0.5]{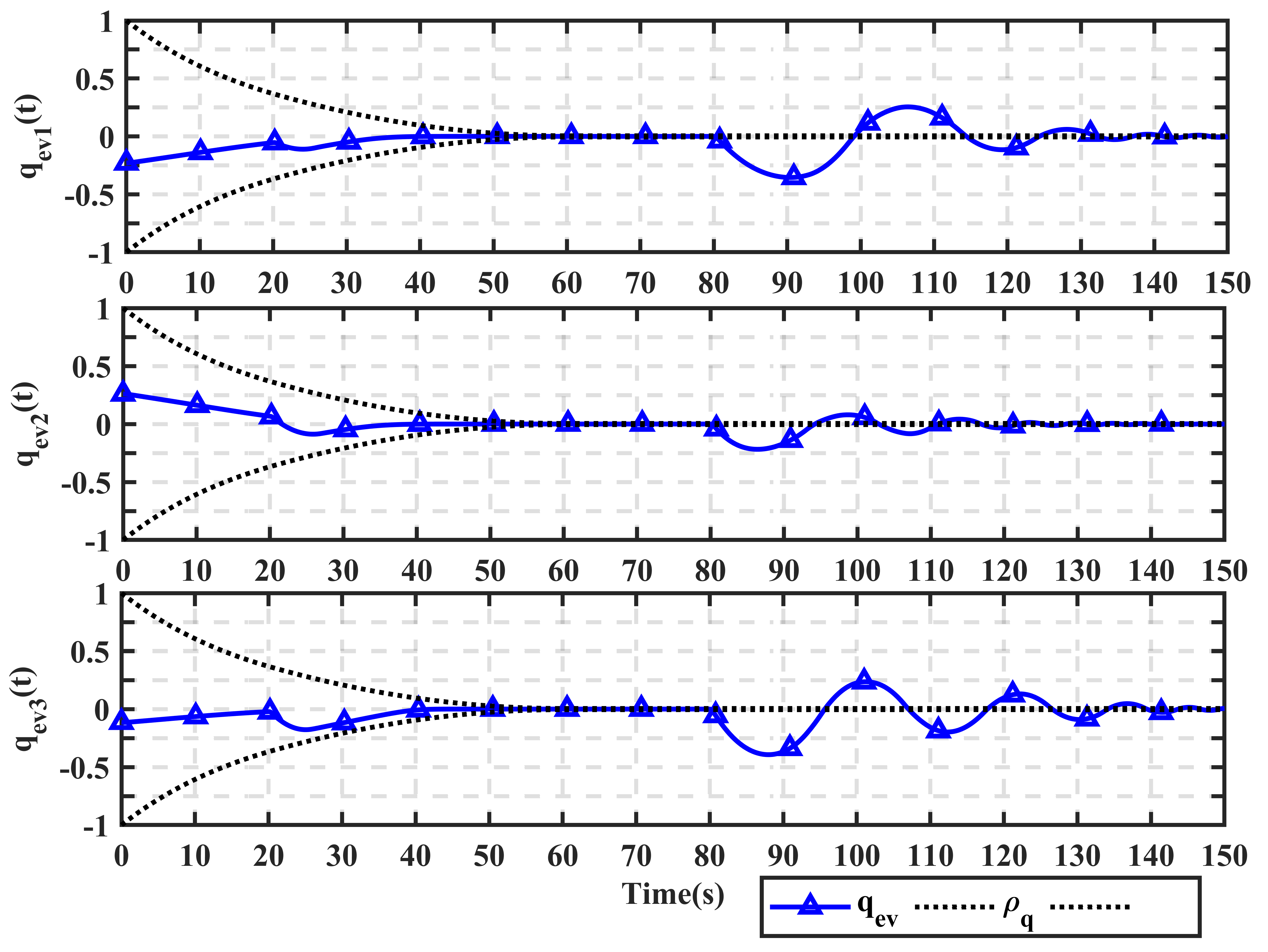}
	\caption{Time Responding of $\boldsymbol{q}_{ev}$ Without Adaptive PPF (Sudden Disturbance Case)} 
	\label{fig_qe_dis_noappf}    
\end{figure}

\begin{figure}[hbt]   
	\centering 
	\includegraphics[scale = 0.5]{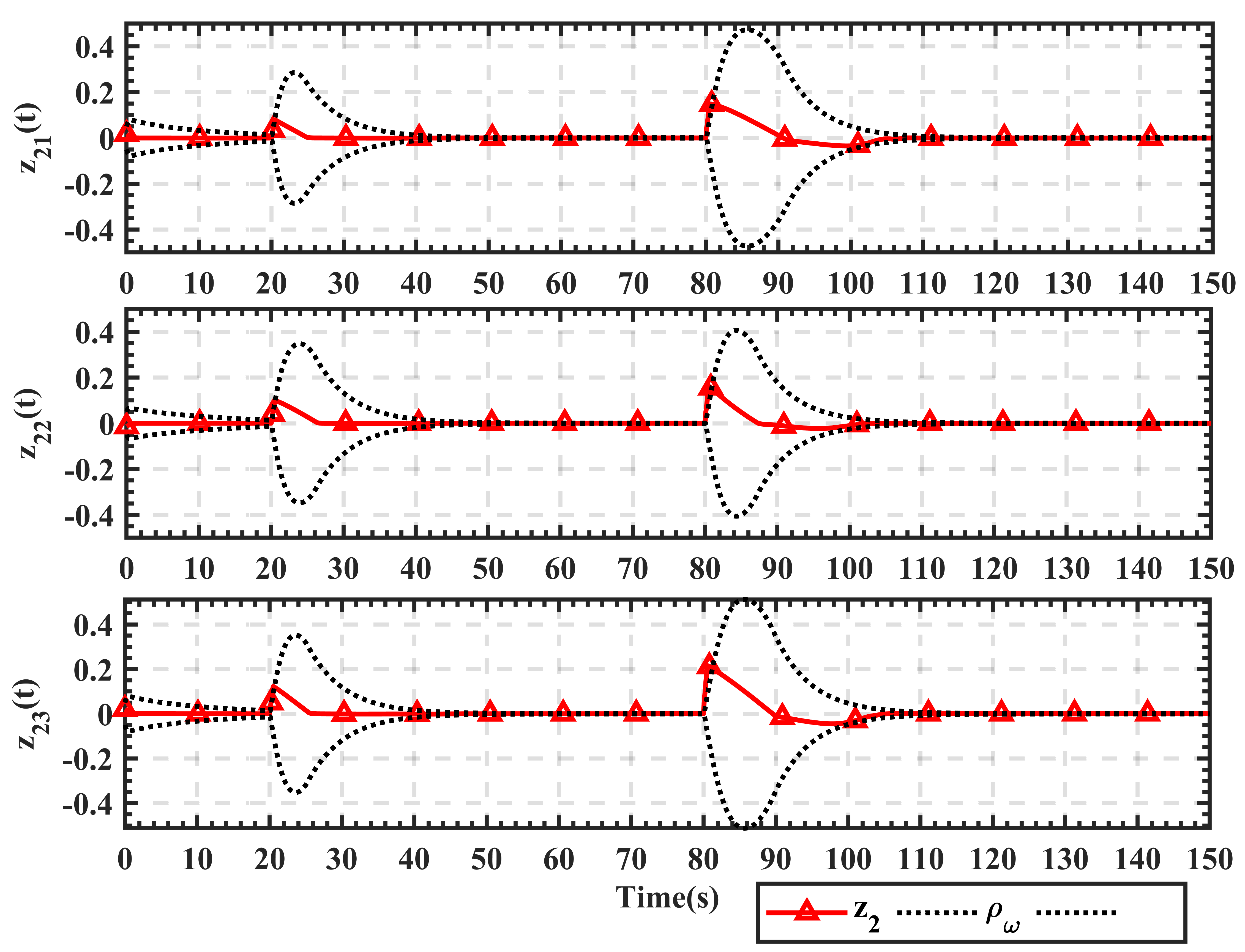}
	\caption{Time Responding of $\boldsymbol{z}_{2}$ (Sudden Disturbance Case)}  
	\label{fig_z_dis}    
	\centering 
	\includegraphics[scale = 0.5]{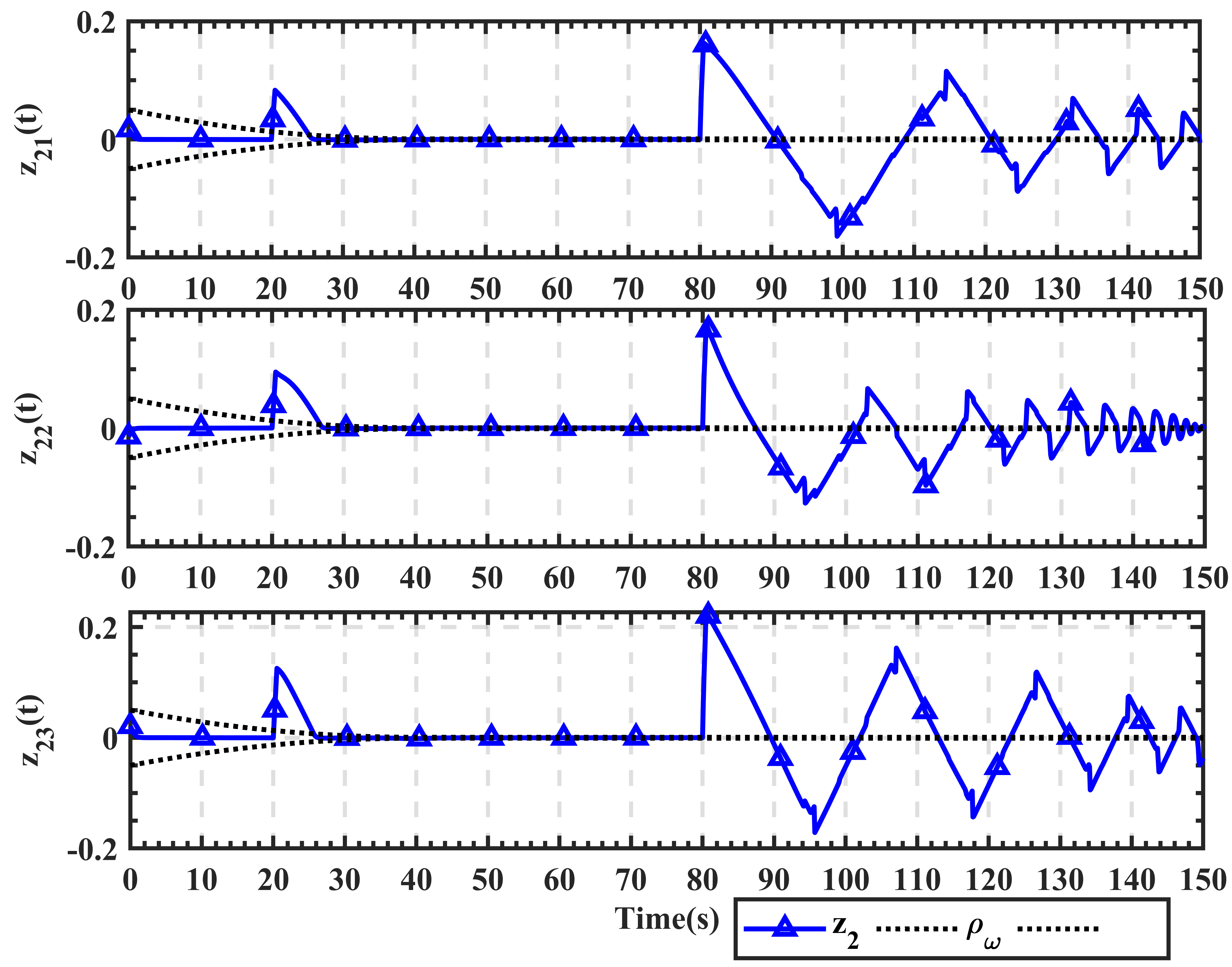}
	\caption{Time Responding of $\boldsymbol{z}_{2}$ Without Adaptive PPF (Sudden Disturbance Case)}  
	\label{fig_z_dis_noappf}     
\end{figure}

The exerted sudden disturbance is modeled as follows: for $t = 20s$, $\boldsymbol{d}_{a} = \left[0.5,0.5,0.5\right]N\cdot m$; for $t = 80s$, $\boldsymbol{d}_{a} = \left[0.8,0.8,0.8\right]N\cdot m$. From Figure [\ref{fig_qe_dis}][\ref{fig_z_dis}], it can be found that although the sudden severe disturbance critically perturbed the system, the proposed scheme is still able to recover in a short time. Also, the transient behavior of the re-stabilizing process is smooth and almost non-overshoot.

 Here we make a simple comparison: we remove the adaptive strategy in the following simulation, which is illustrated in Figure[\ref{fig_qe_dis_noappf}][\ref{fig_z_dis_noappf}]. It can be found that the recover time of the controller without adaptive PPF is much more longer than the other. Since the $\boldsymbol{z}_{2}$ is the outer loop which is directly influenced by the disturbance, it can be observed that without the adaptive for PPF, the $\boldsymbol{z}_{2}$ will chattering for a long time. This influences the tracking to the virtual control law, leading to the chattering of $\boldsymbol{q}_{e}$ finally.

\subsection{Comparison to Traditional PPC} 

In this section, a BLF-based PPC benchmark controller is taken into considering to refer to as a comparison in the evaluation of the singularity circumvent effect. The benchmark controller is denoted as $TradBLF$ in the subsection, and the detailed idea of the benchmark controller can be found in \cite{hu_adaptive_2018-1}, which is designed based on a single-layer PPC structure. Here we omit it for brevity. In the comparison simulation, a sudden disturbances $\boldsymbol{d}_{a} = \left[0.8;0.8;0.8\right]^{\text{T}}$ will be exerted to the system at $t = 80s$.

\begin{figure}[hbt]    
	\centering 
	\includegraphics[scale = 0.55]{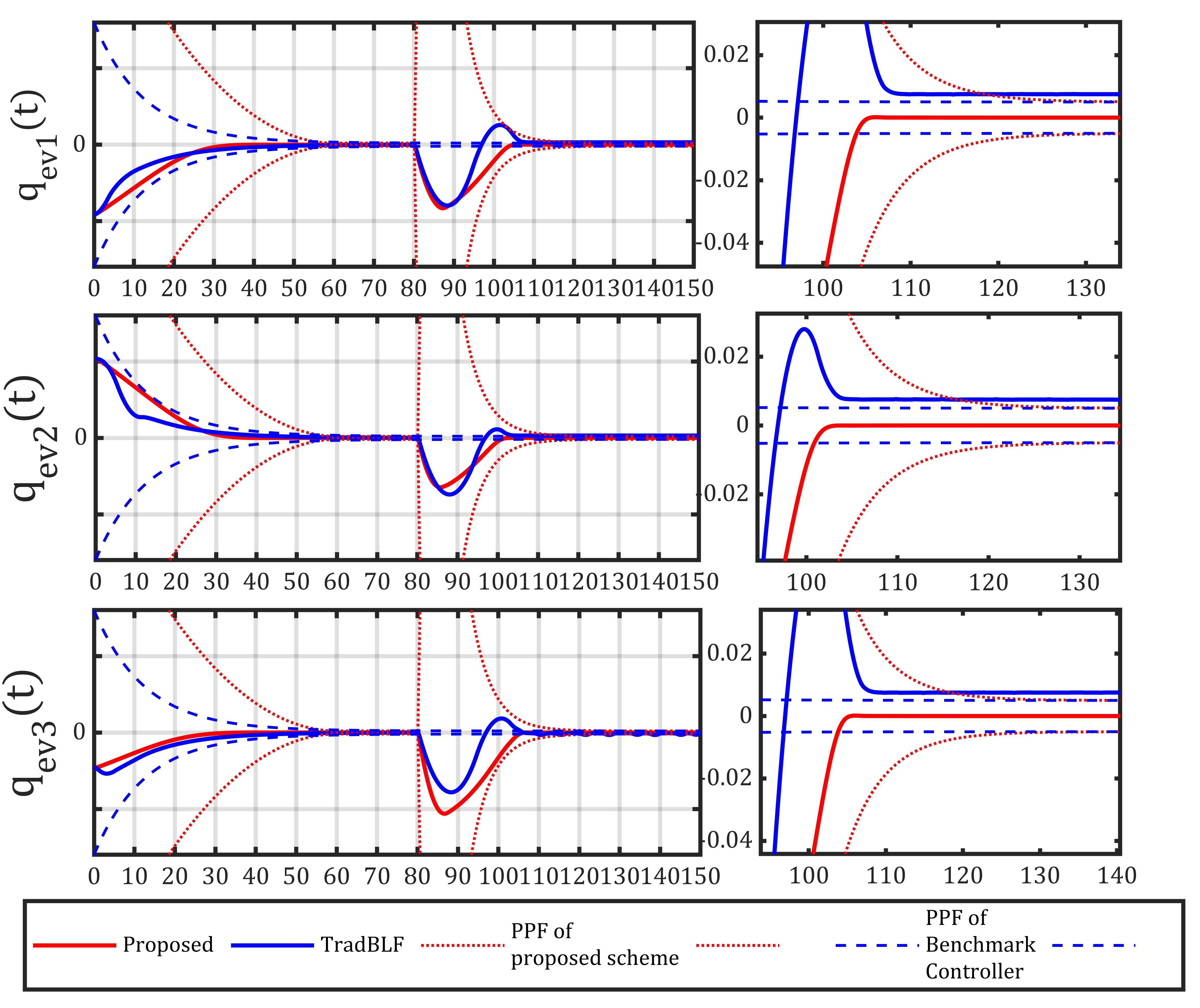}
	\caption{Comparison of the proposed scheme and Benchmark Controller $\boldsymbol{q}_{ev}$}  
	\label{fig_qe_comp} 
	\centering 
	\includegraphics[scale = 0.45]{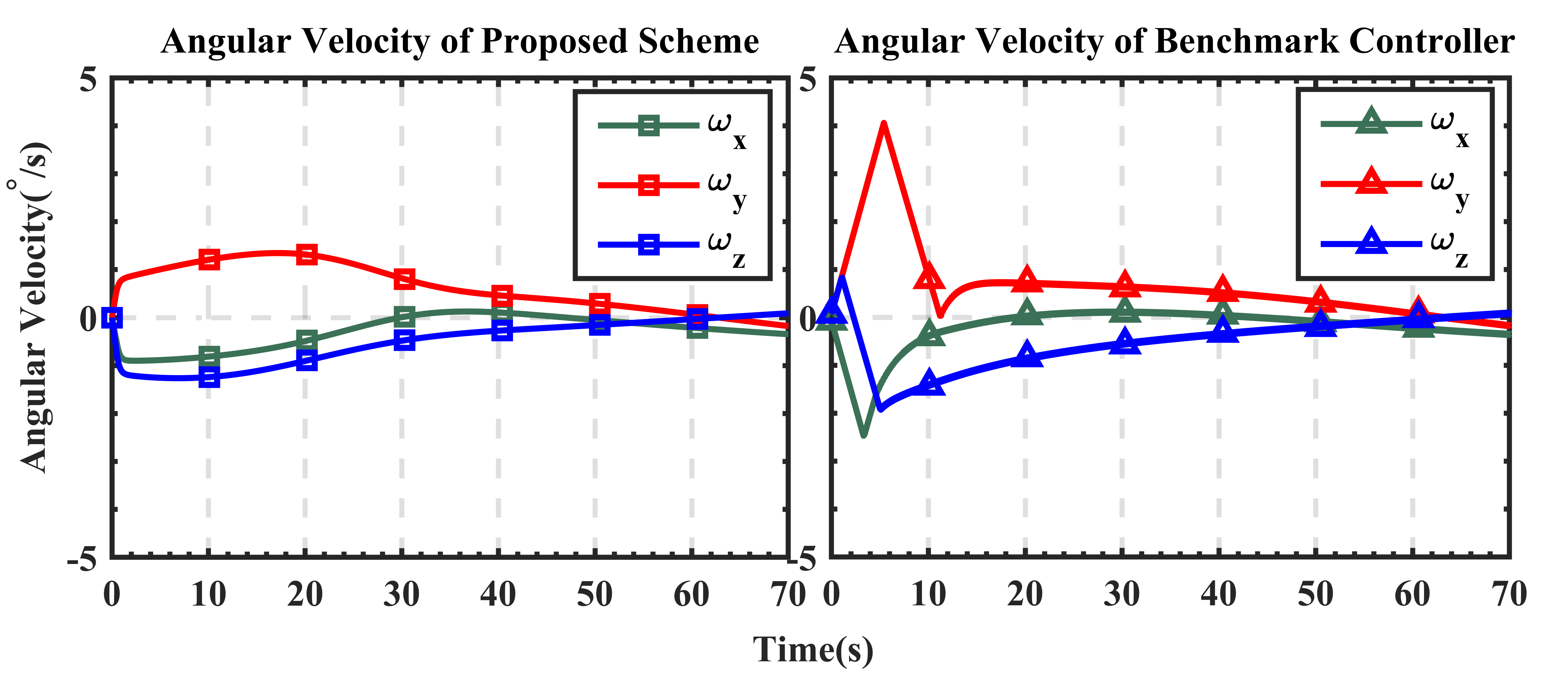}
	\caption{Comparison of the proposed scheme and Benchmark Controller $\boldsymbol{\omega}_{s}$}  
	\label{fig_w_comp}
\end{figure}

The simulation result of the benchmark controller and the proposed scheme is illustrated in Figure [\ref{fig_qe_comp}]. The blue line denotes the state trajectory of the benchmark controller, while the blue dotted line represents its corresponding PPF envelope. The state trajectory of the proposed scheme is illustrated in red line, while the red dotted line stands for its PPF.

When $t = 80s$, both two state trajectory is severely disturbed by the sudden disturbance, as we can find in Figure [\ref{fig_qe_comp}]. Notably, although the state trajectory of the benchmark controller is able to converge, it is failed to converge back to the constraint region. This phenomenon can be explained by the singularity of BLF-type PPC scheme, as we stated in \cite{SAPPC2022}. The proposed scheme however, is able to stabilizing the system into the given region, guaranteeing the performance requirements to be satisfied. 
Further, the maximum angular velocity of $TradPPC$ is much more bigger than the proposed scheme, as illustrated in Figure [\ref{fig_w_comp}].

\section{Conclusion}
This paper focuses on the space attitude tracking problem under physical limitations and performance requirements, with robustness in a critical position during the controller design. For the robustness issue, the inherent singularity problem is tackled by the newly-designed BLF. Owing to the proposed adaptive strategy for PPF, the chattering problem is significantly alleviated, which provides smooth convergence of the system after severe disturbances. Subsequently, the angular velocity constraint is satisfied by applying the proposed BLF, and the input saturation issue is solved by introducing the auxiliary system. Based on the designed BLF, we derive a backstepping controller with a double-layer PPC structure, ensuring that the performance requirements can be satisfied. The numerical simulation results validate our theoretical analysis, providing convincing support for the proposed scheme. Further research in this issue will focus on other more complex constraints, such as the pointing constraints, which are hard to fuse with the performance requirements up to now and worth further investigation.

\bibliographystyle{IEEEtran}
\bibliography{AIAA_ppccons_2}

\begin{thebibliography}{10}
\providecommand{\url}[1]{#1}
\csname url@samestyle\endcsname
\providecommand{\newblock}{\relax}
\providecommand{\bibinfo}[2]{#2}
\providecommand{\BIBentrySTDinterwordspacing}{\spaceskip=0pt\relax}
\providecommand{\BIBentryALTinterwordstretchfactor}{4}
\providecommand{\BIBentryALTinterwordspacing}{\spaceskip=\fontdimen2\font plus
\BIBentryALTinterwordstretchfactor\fontdimen3\font minus
  \fontdimen4\font\relax}
\providecommand{\BIBforeignlanguage}[2]{{%
\expandafter\ifx\csname l@#1\endcsname\relax
\typeout{** WARNING: IEEEtran.bst: No hyphenation pattern has been}%
\typeout{** loaded for the language `#1'. Using the pattern for}%
\typeout{** the default language instead.}%
\else
\language=\csname l@#1\endcsname
\fi
#2}}
\providecommand{\BIBdecl}{\relax}
\BIBdecl

\bibitem{wie1995feedback}
\BIBentryALTinterwordspacing
B.~Wie and J.~Lu, ``Feedback control logic for spacecraft eigenaxis rotations
  under slew rate and control constraints,'' \emph{Journal of Guidance,
  Control, and Dynamics}, vol.~18, no.~6, pp. 1372--1379, 1995. [Online].
  Available: \url{https://doi.org/10.2514/3.21555}
\BIBentrySTDinterwordspacing

\bibitem{hu2013robust}
\BIBentryALTinterwordspacing
Q.~Hu, B.~Li, and Y.~Zhang, ``Robust attitude control design for spacecraft
  under assigned velocity and control constraints,'' \emph{ISA transactions},
  vol.~52, no.~4, pp. 480--493, 2013. [Online]. Available:
  \url{https://doi.org/10.1016/j.isatra.2013.03.003}
\BIBentrySTDinterwordspacing

\bibitem{li2016adaptive}
\BIBentryALTinterwordspacing
M.~Li, M.~Hou, and C.~Yin, ``Adaptive attitude stabilization control design for
  spacecraft under physical limitations,'' \emph{Journal of guidance, control,
  and dynamics}, vol.~39, no.~9, pp. 2179--2183, 2016. [Online]. Available:
  \url{https://doi.org/10.2514/1.G000348}
\BIBentrySTDinterwordspacing

\bibitem{shao2018fault}
\BIBentryALTinterwordspacing
X.~Shao, Q.~Hu, Y.~Shi, and B.~Jiang, ``Fault-tolerant prescribed performance
  attitude tracking control for spacecraft under input saturation,'' \emph{IEEE
  Transactions on Control Systems Technology}, vol.~28, no.~2, pp. 574--582,
  2018. [Online]. Available: \url{https://doi.org/10.1109/TCST.2018.2875426}
\BIBentrySTDinterwordspacing

\bibitem{sun2017disturbance}
\BIBentryALTinterwordspacing
L.~Sun and Z.~Zheng, ``Disturbance-observer-based robust backstepping attitude
  stabilization of spacecraft under input saturation and measurement
  uncertainty,'' \emph{IEEE Transactions on Industrial Electronics}, vol.~64,
  no.~10, pp. 7994--8002, 2017. [Online]. Available:
  \url{https://doi.org/10.1109/TIE.2017.2694349}
\BIBentrySTDinterwordspacing

\bibitem{zou2016finite}
\BIBentryALTinterwordspacing
A.-M. Zou, A.~H. de~Ruiter, and K.~D. Kumar, ``Finite-time output feedback
  attitude control for rigid spacecraft under control input saturation,''
  \emph{Journal of the Franklin Institute}, vol. 353, no.~17, pp. 4442--4470,
  2016. [Online]. Available:
  \url{https://doi.org/10.1016/j.jfranklin.2016.08.013}
\BIBentrySTDinterwordspacing

\bibitem{hu2018adaptive}
\BIBentryALTinterwordspacing
Q.~Hu, Y.~Shi, and X.~Shao, ``Adaptive fault-tolerant attitude control for
  satellite reorientation under input saturation,'' \emph{Aerospace Science and
  Technology}, vol.~78, pp. 171--182, 2018. [Online]. Available:
  \url{https://doi.org/10.1016/j.ast.2018.04.015}
\BIBentrySTDinterwordspacing

\bibitem{zhang2019observer}
\BIBentryALTinterwordspacing
C.~Zhang, G.~Ma, Y.~Sun, and C.~Li, ``Observer-based prescribed performance
  attitude control for flexible spacecraft with actuator saturation,''
  \emph{ISA transactions}, vol.~89, pp. 84--95, 2019. [Online]. Available:
  \url{https://doi.org/10.1016/j.isatra.2018.12.027}
\BIBentrySTDinterwordspacing

\bibitem{wei2018learning}
\BIBentryALTinterwordspacing
C.~Wei, J.~Luo, H.~Dai, and G.~Duan, ``Learning-based adaptive attitude control
  of spacecraft formation with guaranteed prescribed performance,'' \emph{IEEE
  transactions on cybernetics}, vol.~49, no.~11, pp. 4004--4016, 2018.
  [Online]. Available: \url{https://doi.org/10.1109/TCYB.2018.2857400}
\BIBentrySTDinterwordspacing

\bibitem{wei2021overview}
\BIBentryALTinterwordspacing
C.~Wei, Q.~Chen, J.~Liu, Z.~Yin, and J.~Luo, ``An overview of prescribed
  performance control and its application to spacecraft attitude system,''
  \emph{Proceedings of the Institution of Mechanical Engineers, Part I: Journal
  of Systems and Control Engineering}, vol. 235, no.~4, pp. 435--447, 2021.
  [Online]. Available: \url{https://doi.org/10.1177/0959651820952552}
\BIBentrySTDinterwordspacing

\bibitem{bechlioulis_adaptive_2009}
\BIBentryALTinterwordspacing
C.~P. Bechlioulis and G.~A. Rovithakis, ``Adaptive control with guaranteed
  transient and steady state tracking error bounds for strict feedback
  systems,'' \emph{Automatica}, vol.~45, no.~2, pp. 532--538, 2009. [Online].
  Available: \url{https://doi.org/10.1016/j.automatica.2008.08.012}
\BIBentrySTDinterwordspacing

\bibitem{SAPPC2022}
\BIBentryALTinterwordspacing
J.~Lei, T.~Meng, W.~Wang, H.~Li, and Z.~Jin, ``Singularity-avoidance prescribed
  performance attitude tracking of spacecraft,'' 2022. [Online]. Available:
  \url{https://doi.org/10.48550/arXiv.2206.12761}
\BIBentrySTDinterwordspacing

\bibitem{yong2020flexible}
\BIBentryALTinterwordspacing
K.~Yong, M.~Chen, Y.~Shi, and Q.~Wu, ``Flexible performance-based robust
  control for a class of nonlinear systems with input saturation,''
  \emph{Automatica}, vol. 122, 2020. [Online]. Available:
  \url{https://doi.org/10.1016/j.automatica.2020.109268}
\BIBentrySTDinterwordspacing

\bibitem{WANG2022}
\BIBentryALTinterwordspacing
K.~Wang, T.~Meng, W.~Wang, R.~Song, and Z.~Jin, ``Finite-time extended state
  observer based prescribed performance fault tolerance control for spacecraft
  proximity operations,'' \emph{Advances In Space Research}, 2022. [Online].
  Available: \url{https://doi.org/10.1016/j.asr.2022.05.072}
\BIBentrySTDinterwordspacing

\bibitem{golestani2022prescribed}
\BIBentryALTinterwordspacing
M.~Golestani, S.~Mobayen, S.~U. Din, F.~F. El-Sousy, M.~T. Vu, and
  W.~Assawinchaichote, ``Prescribed performance attitude stabilization of a
  rigid body under physical limitations,'' \emph{IEEE Transactions on Aerospace
  and Electronic Systems}, 2022. [Online]. Available:
  \url{https://doi.org/10.1109/TAES.2022.3158371}
\BIBentrySTDinterwordspacing

\bibitem{xiao2011adaptive}
\BIBentryALTinterwordspacing
B.~Xiao, Q.~Hu, and Y.~Zhang, ``Adaptive sliding mode fault tolerant attitude
  tracking control for flexible spacecraft under actuator saturation,''
  \emph{IEEE Transactions on Control Systems Technology}, vol.~20, no.~6, pp.
  1605--1612, 2011. [Online]. Available:
  \url{https://doi.org/10.1109/TCST.2011.2169796}
\BIBentrySTDinterwordspacing

\bibitem{bechlioulis_robust_2008}
\BIBentryALTinterwordspacing
C.~P. Bechlioulis and G.~A. Rovithakis, ``Robust adaptive control of feedback
  linearizable mimo nonlinear systems with prescribed performance,'' \emph{IEEE
  Transactions on Automatic Control}, vol.~53, no.~9, pp. 2090--2099, 2008.
  [Online]. Available: \url{https://doi.org/10.1109/TAC.2008.929402}
\BIBentrySTDinterwordspacing

\bibitem{hu_adaptive_2018-1}
\BIBentryALTinterwordspacing
Q.~Hu, Y.~Shi, and X.~Shao, ``Adaptive fault-tolerant attitude control for
  satellite reorientation under input saturation,'' \emph{Aerospace Science and
  Technology}, vol.~78, pp. 171--182, 2018. [Online]. Available:
  \url{https://doi.org/10.1016/j.ast.2018.04.015}
\BIBentrySTDinterwordspacing

\end{thebibliography}

\end{document}